\documentclass[final,5p,times,twocolumn,authoryear]{elsarticle}

\usepackage{amssymb}
\usepackage{float}
\usepackage{lipsum}
 \usepackage{xcolor} 
\usepackage{float} % in preamble
\usepackage{orcidlink}
\usepackage[normalem]{ulem} 
\usepackage{amsmath}
\usepackage[normalem]{ulem}
\usepackage{multirow}
\usepackage{array}
\usepackage{url}
\usepackage[caption=false]{subfig} 
\usepackage[varvw]{newtxmath}
\usepackage{adjustbox}

\journal{High Energy Astrophysics}

\begin{document}

\begin{frontmatter}

\title{Probing TeV Afterglow Emission of GRB~221009A with Gaussian Structured jet in Wind-driven medium}

\author[first]{Tanima Mondal\orcidlink{0000-0002-9445-1405}}
\affiliation[first]{organization={Department of Physics, Indian Institute of Technology Kharagpur},%Department and Organization
            addressline={}, 
            city={Kharagpur},
            postcode={721302}, 
            state={West Bengal},
            country={India}}

\author[second]{Sabyasachi Chakraborty\orcidlink{0009-0001-8084-0565}}
\affiliation[second]{organization={Department of Physics, Indian Institute of Technology Kharagpur},%Department and Organization
            addressline={}, 
            city={Kharagpur},
            postcode={721302}, 
            state={West Bengal},
            country={India}}

\author[third]{Lekshmi Resmi\orcidlink{0000-0001-9407-9845}}
\affiliation[third]{organization={Department of Earth \& Space Sciences, Indian Institute of Space Science \& Technology},%Department and Organization
            addressline={}, 
            city={Trivandrum},
            postcode={695547}, 
            state={Kerala},
            country={India}}

\author[fourth]{Debanjan Bose\orcidlink{0000-0003-1071-5854}}
\affiliation[third]{organization={Department of Physics, Central University of Kashmir},%Department and Organization
            addressline={}, 
            city={Ganderbal},
            postcode={191131}, 
            state={Jammu \& Kashmir},
            country={India}}

\begin{abstract}

Recent detections of very high energy (VHE; GeV–TeV) photons from gamma-ray burst (GRB) afterglows, most notably the extreme event GRB 221009A, demand refined theoretical frameworks that incorporate both realistic jet structures and complex circumburst environments. The angular structure of GRB jets plays a crucial role in determining afterglow signatures. In one of our recent studies we have seen, Gaussian jets, with their smoother angular decline, naturally produce early and bright afterglow peaks for on-axis observers and delayed, softer, and dimmer peaks at higher inclinations. This shallow decline or slowly brightening afterglow, resulting from a Gaussian jet structure, further suppresses excessive lateral expansion, unlike the abrupt jet edge found in top-hat models. Thus, Gaussian jets become one of the compelling alternatives to other structured jet and top-hat models for explaining afterglow behaviour. In this study, we have implemented a Gaussian structured jet model that primarily explains TeV afterglow behaviour from structured jets propagating as adiabatic forward shocks in a wind-driven medium. This study shows that the peak time and flux of TeV afterglow emission strongly depend on jet geometry, total energy of the jet and circumburst wind density, with the microphysical parameters ratio that scales the SSC flux profile. We showcase the VHE detection of sub-TeV photons, illustrating the favourable afterglow parameter regime with the Cherenkov Telescope Array (CTA). We find out of all simulated TeV events, only $\sim10\%$ of them exceed the CTA sensitivity in wind-driven medium, favouring near core-aligned views, with high kinetic energy and wind density, moderate initial Lorentz factor and downstream magnetic field, and higher value of fractional thermal energy density transferred to the non-thermal electrons. We apply the VHE Gaussian structured jet model to GRB 221009A, performing multi-band fits that include wind-modified dynamics, Klein-Nishina suppression, and EBL attenuation. The fits favour a mildly off-axis geometry and successfully reproduce the observed X-ray, GeV-TeV light curves.
\end{abstract}


\end{frontmatter}

\section{Introduction}
\label{introduction}

Gamma-ray bursts (GRBs) are among the most luminous and energetic explosions in the Universe, releasing enormous amounts of energy across the electromagnetic spectrum \citep{Costa_1997Natur.783C, Kulkarni:1999aa}. Their broadband afterglow emission, from radio to very high energy (VHE) $\gamma$ rays, is widely interpreted as synchrotron radiation from relativistic electrons accelerated in external forward shocks~\citep{optical_prediction, Costa_1997Natur.783C, Van1997Natur_686V, frail1997Natur_261F, Sari1998ApJ_17S, Llyod2000ApJ.722L, burgess2020NatAs.174B}. Over the past few years, detections of GeV–TeV photons by ground based Cherenkov telescopes have challenged this simple single–zone synchrotron picture~\citep{acciari2019magic, abdalla2019very, hess2021revealing, lhaaso2023very, abe2024magic}. A leading explanation for this additional high energy component is synchrotron self–Compton (SSC) emission, in which the same relativistic electrons that produce the synchrotron photons upscatter them to the TeV band~\citep{chiang1999synchrotron, dermer2000beaming, sari2001synchrotron, murase2011implications, fraija2019synchrotron, wang2019synchrotron, derishev2019physical, joshi2021modelling, mondal2023probing, mondal2025follow,ren2024jet}. These observations collectively point toward the need for refined theoretical models that self-consistently capture the microphysics of particle acceleration, radiative cooling, and multi–zone emission processes.

The discovery of GRB~221009A has provided an unprecedented opportunity to explore such extreme regimes. Triggered by \textit{Fermi}/GBM\footnote{\url{https://fermi.gsfc.nasa.gov/science/instruments/gbm.html}} on 2022 October 9 at 13:16:59~UT ($T_0$) and located at a relatively low redshift ($z \!\sim\! 0.151$)~\citep{deugarte2022_1D, veres2022grb}, it stands among the brightest and most energetic long GRBs ever detected. The burst was extensively followed throughout the electromagnetic spectrum by \textit{Swift}~\citep{williams2023grb}, \textit{Fermi}–LAT~\citep{lesage2023fermi}, INTEGRAL/SPI–ACS, and Konus–Wind~\citep{frederiks2023properties}, with its localisation refined through InterPlanetary Network triangulation~\citep{2022GCN.32641....1S}. Most remarkably, LHAASO detected photons exceeding $10$~TeV~\citep{lhaaso2023very, huang2022lhaaso}, marking the first ultra high energy (UHE) $\gamma$ ray observation from a long GRB. This detection firmly establishes GRB~221009A as a benchmark event for testing particle acceleration and radiation mechanisms in relativistic outflows, and necessitates the joint multiwavelength and multimessenger observations in the VHE regime.

Several studies have modeled multi-band afterglows with both top-hat and structured jet outflows (power-law, two-component) to explain the characteristics of peak times, peak fluxes, and viewing-angle effects \citep{mooley2018mildly, margutti2017electromagnetic, ren2023possibility, o2023structured, sato2023two, gill2023grb, ren2024jet}. Among these, narrow top-hat jets struggle to reproduce TeV afterglows, as their strong lateral spreading leads to post-peak declines that are too steep and late-time fluxes that are too faint, restricting fits to the earliest epochs \citep{margutti2017electromagnetic, mooley2018mildly}. In wind environments, the sharp-edged, uniform core of a top-hat leads to unrealistically large isotropic energies to match the observed brightness. For wind environments, seminal studies established the baseline synchrotron and SSC emission and dynamical evolution for uniform jets \citep{chevalier2000wind, panaitescu2000analytic}. More recently, structured outflows have been shown to account for the VHE afterglow of GRB~221009A, where \citet{gill2023grb} invoke a shallow angular structure, while \citet{Ren_2024} employ a top-hat inner
core embedded in a power-law structured outer wing, and introduce an “inverted” configuration to capture ISM-to-wind transitions.

Motivated by these developments, we therefore adopt a Gaussian structured jet evolving in an adiabatic forward-shock wind-driven medium. For the Gaussian jet structure, the observed flux decreases less steeply and follows a smooth distribution, showcasing a significant non-zero emission profile even for observers aligning far away from the jet core. In comparison with a Gaussian structure, the shallow power–law jet structure~\citep{gill2023grb, o2023structured} ($\epsilon(\theta)\propto\theta^{-a}$ with $a\!\lesssim\!2$) suffers from certain issues. The total energy in such a power-law configuration increases too slowly toward the wings for $a\!\le\!2$, making it necessary to introduce an outer cutoff angle $\theta_{\max}$ to keep the total energy within realistic limits~\citep{zhang2002gamma}. Further, the shallow wings keep substantial energy at large $\theta$. This leads to significant over-brightening of the off-axis or late-time emission, which results in a flattening of the post-peak decay of the afterglow \citep{granot2003constraining}. Furthermore, the power-law wings cause a strong coupling between the viewing angle $\theta_v$ and the outer cutoff $\theta_{\max}$. Because of this, the parameters such as the core angular size $\theta_c$ and power-law index $a$ are difficult to determine precisely from observations~\citep{granot2003constraining,o2023structured}. 

A Gaussian jet structure naturally reproduces the following observed behaviour, where early and bright peaks occur for observers aligned near the jet core, while misaligned observers see delayed and dimmer peaks \citep{mondal2025follow}. This behaviour reflects the angular energy profile and relativistic beaming effects inherent in Gaussian-structured jets. In the TeV energy range where synchrotron self-Compton (SSC) emission dominates, the Gaussian jet structure enhances the seed synchrotron photon field near the core. This configuration avoids excessive emission from the jet wings, resulting in higher peak fluxes without the need for extreme microphysical parameters. Additionally, the Gaussian profile remains stable against the rapid lateral expansion that challenges top-hat jet models in wind environments. This Gaussian model provides better control over relativistic beaming effects and consistently reproduces both temporal and spectral evolution for on-axis and off-axis observers~\citep{resmi2018low, mondal2025follow}. Thus, a smooth Gaussian core and its wings propagating in a wind-like medium with density profile $\rho \propto r^{-2}$ become particularly critical at TeV energies. In this regime, even small variations in jet geometry and afterglow parameters can drastically alter the observed brightness by orders of magnitude. This sensitivity becomes especially evident when interpreting TeV observations with the Cherenkov Telescope Array (CTA), where slight changes in viewing angle, microphysics, and wind density parameter significantly impact detectability.

To address these challenges, we develop an afterglow model for a Gaussian structured jet evolving in a wind-driven medium and explore the VHE afterglow detection prospects with CTA for both on- and off-axis observers. Applying this framework to the Xray-GeV-TeV data of GRB 221009A, we perform a multi-band fit using Markov Chain Monte Carlo (MCMC) methods to robustly constrain the model parameters. While long-duration GRB progenitors naturally favour wind environments, Gaussian angular profiles provide a soft-edged, energetically efficient structure. However, the combined wind-Gaussian scenario remains comparatively underexplored in the TeV domain. Our model integrates jet geometry (parameterised by the viewing-to-core angle ratio $\theta_v/\theta_c$), dynamics (accounting for wind-modified deceleration), and high-energy radiation processes (including synchrotron self-Compton with Klein-Nishina suppression, $\gamma\gamma$ absorption, and extragalactic background light ---EBL attenuation). Given the strong dependence of TeV light curves on these factors, it is vital to systematically vary the parameter space to understand their influence on the afterglow peak time and peak flux.

The paper is organised as follows. In Section~\ref{sec:model}, we present the off-axis Gaussian structured-jet afterglow model in a wind-driven environment and outline its key features. Section~\ref{sec:param_charc} examines how the afterglow model parameters shape the TeV light curve. In Section~\ref{sec:detect_TeV_events} we explore the detectability of TeV afterglows in a wind environment relative to the sensitivity of the Cherenkov Telescope Array (CTA). To validate the framework, in Section~\ref{sec:fit_VHE_data}, we apply our Gaussian structured-jet afterglow model to GRB~221009A, jointly fitting its X-ray, GeV, and TeV data in a wind medium. Finally, we summarie our results in Section~\ref{sec:conclusions}.

\section{Afterglow emission from structured jet in stellar wind driven medium} \label{sec:model}

In this study, we develop an analytical framework to model the broadband afterglow emission from a Gaussian structured jet interacting with a stellar wind environment. For structured jet decelerating in a homogeneous medium, we assume a jet decelerating into a cold stellar wind driven medium where its power-law mass density profile becomes $\rho(r)=Ar^{-2}= (5 \times 10^{11} A_{\star})\times r^{-2}  \: gm \: cm^{-1}$. $A$ is defined by the ratio of mass-loss rate ($\dot{M}$) of the progenitor Wolf-Rayet star to the wind velocity ($v_{w}$). Here we consider the typical $\dot{M} = 10^{-5}\, M_{\odot}\, \mathrm{yr}^{-1}$ and $v_w = 10^8\, \mathrm{cm\,s}^{-1}$ for Wolf-Royet star \citep{chevalier1999gamma, chevalier2000wind}. $A_{\star}$ is the dimensionless normalisation parameter that determines the density profile of a stellar medium. Hence, the radial number density of the wind becomes, %$n(r)=\frac{A}{m_p}\,r^{-2} = \frac{5 \times 10^{11} A_{\star}}{m_p} r^{-2} = n_{k} r^{-2}\;\mathrm{cm^{-3}}$, 
\begin{align}\label{eq:wind_density}
\begin{split}
        n(r)=\frac{A}{m_p}\,r^{-2} &= \frac{5 \times 10^{11} A_{\star}}{m_p} r^{-2}\\ 
        &= n_{k} r^{-2}\;\mathrm{cm^{-3}}
\end{split}
\end{align}
where $n_{k} \sim (3 \times 10^{35})A_{\star}$ is the wind density normalisation factor, $m_p$ is the mass of a proton, and r represents the radial distance from the central engine of the GRB. In this framework, the relativistic Gaussian jet decelerates as it interacts with the stellar wind medium, driving a forward shock that accelerates electrons and amplifies magnetic fields. This interaction leads to broadband afterglow emission through synchrotron and SSC emission processes extending from radio to TeV energies.

\subsection{Jet Structure and Circumburst Environment}
\label{sec:dynamics}

In the stellar wind-driven medium, to model the evolution of a structured jet, we retain the intrinsic Gaussian angular structure $\{\mathcal{E}(\theta), \Gamma_0(\theta)\beta_0(\theta)\}$ identical to that used in \citet{resmi2018low} and \citet{mondal2025follow}. Both the energy and Lorentz factor vary with polar angle and follow a Gaussian profile \citep{lamb2017electromagnetic, resmi2018low}:

\begin{equation}\label{eq:gauss_profile}
\mathcal{E}(\theta) = \mathcal{E}_c \exp\left(-\frac{\theta^2}{2\theta_c^2}\right), \quad
\Gamma_0(\theta)\beta_{0}(\theta) = \Gamma_c \beta_{c} \exp\left(-\frac{\theta^2}{2\theta_c^2}\right).
\end{equation}

$\mathcal{E}(\theta)$ represents the kinetic energy per unit solid angle, characterised by kinetic energy at jet core $\mathcal{E}_c = E_{k}/\pi \theta_{c}^{2}(1-e^{-\theta_{j}^{2}/\theta_{c}^2})$, with $E_{k}$ being the total kinetic energy of the jet. The characteristic angle $\theta_{c}$ denotes the jet core width, which determines the angular steepness of the jet structure. The initial velocity distribution of the jet is expressed through $\Gamma_0(\theta)\beta_{0}(\theta)$, which follows the same Gaussian angular dependence.

In the wind-driven medium, only the radial evolution is modified according to the wind density profile $n(r) \propto r^{-2}$. The inclination angle of a surface element at $(\theta_i, \phi_k)$, as observed from a viewing angle $\theta_v$, remains analogous to the ISM case \citep{mondal2025follow} and is given by, 

\begin{equation}
    \cos \alpha_{\rm inc}^{i,k} = \cos\theta_i\cos\theta_v + \sin\theta_i\sin\theta_v\cos\phi_k,
\end{equation}
where the polar angle $\theta_i$ spans $0 < \theta_i < \pi/2$ and the azimuthal angle $\phi_k$ ranges over $0 < \phi_k < 2\pi$ with respect to the jet axis~\citep{resmi2018low, mondal2025follow}. For each angular element $(i,k)$, we consider the adiabatic evolution of the bulk motion of the ultra-relativistic ejecta as it propagates through the wind-driven medium as,

\begin{equation}
\label{eq:Gb_wind}
    \Gamma(\theta,r)\,\beta(\theta,r)=\Gamma_0(\theta)\,\beta_0(\theta)
    \left(\frac{r}{r_{\rm dec}}\right)^{-1/2}.
\end{equation}

In a stellar-wind environment, the deceleration dynamics differ from the uniform ISM case because the swept-up mass per steradian increases linearly with radius and is given by $m(r)=\int 4\pi r^2 n(r)\,dr =4\pi A r$. Hence, the deceleration radius becomes,

\begin{equation}
\label{eq:rdec_wind}
    r_{\rm dec}^{wind}=\frac{\mathcal{E}_c}{4\pi\,n_{k}\, \text{c}^{2}\,\Gamma_c^{2}\beta_{c}^{2}}.
\end{equation}

Here, $\Gamma_{c}\beta_{c} = \eta_{c}$ denotes the initial jet velocity at jet core, and $\text{c}$ is the speed of light. Since the jet-core kinetic energy per unit solid angle $\mathcal{E}_c$, strongly depends on the jet core angle $\theta_{c}$ and total jet kinetic energy $E_{k}$, the deceleration radius $r_{\rm dec}$ is highly sensitive to these intrinsic jet parameters~\citep{mondal2025multi}.

For a given radial profile, we compute the observer time $ t_{\rm obs}(r,\alpha_{\rm inc}^{i,k}) $ for each inclination angle following \citet{mondal2025follow}, by interpolating the relation  

\begin{equation}
\label{tobs_eqn}
t_{\rm obs}(r,\alpha_{\rm inc}^{i,k}) = \frac{r}{\beta(r)c}
\left[ 1 - \beta(r)\cos\alpha_{\rm inc}^{i,k} \right].
\end{equation}

In the ultra-relativistic limit ($ \Gamma \gg 1 $), we use the approximations $\beta \simeq 1 - \frac{1}{2\Gamma^{2}}$ and  $1 - \beta \cos\alpha \simeq \frac{1}{2\Gamma^{2}} + \frac{\alpha^{2}}{2}$. For a stellar-wind environment, the Lorentz factor evolves as $\Gamma(r) \propto r^{-1/2}$,  which implies $r(t) \propto t^{1/2}$ and consequently $\Gamma(t) \propto t^{-1/4}$\citep{chevalier2000wind}. This slower decline of $\Gamma$ in a wind medium compared to the ISM causes the visibility peak ($\Gamma \sim 1/\theta_{v}$) to occur at a later time for the same viewing offset. 

From equation~\eqref{eq:rdec_wind} we see that the dynamical deceleration radius in a stellar wind scales as $r_{\rm dec}^{\rm wind} \propto \left( \frac{\mathcal{E}_c}{\eta_c^{2}A} \right)$, whereas in uniform ISM this deceleration radius scale as $r_{\rm dec}^{\rm ISM} \propto \left(\frac{\mathcal{E}_c}{\eta_c^{2}n_0}\right)^{1/3}$~\citep{mondal2025follow}. This $r_{dec}^{wind}$ is typically a few orders of magnitude smaller than that in a uniform ISM for standard parameters because of the high initial density of the wind-driven medium. Although this implies an earlier onset of deceleration by radius or time in the wind case, the more gradual decline of $\Gamma$ prevents an earlier off-axis peak. The condition $\Gamma \sim 1/\theta_{v}$ is reached at a later observer time compared to the ISM.

\subsection{Effect of a Stellar-Wind medium on magnetic field strength, peak flux, and peak time}
\label{subsec:wind_effects}

The afterglow peak flux across the broadband spectrum, from low-frequency radio to VHE gamma rays, is primarily governed by synchrotron and SSC emission. We consider an adiabatic forward shock expanding into a stellar wind-driven environment with a density profile $ n(r)$ (equation~\eqref{eq:wind_density}). In the ultra-relativistic regime, the evolution of each angular jet element in the wind-driven medium yields a comoving post-shock magnetic field strength given by,

\begin{equation}
B^{'}(\theta, r) \;=\; \left( 32\pi\,\epsilon_B\,\Gamma(\theta)^{2}\,n(r)\,m_p c^{2} \right)^{1/2}.
\end{equation}

To describe the role of characteristic Lorentz factor, we recall that electrons accelerated at the shock front are injected into a non-thermal power-law energy distribution, where $N_e(\gamma) \propto \gamma^{-p}$ over $\gamma_m \le \gamma \le \gamma_{\max}$, where $p\gtrsim 2$. The minimum Lorentz factor of the injected electrons is given $\gamma_m(\theta) = g(p)\epsilon_e(m_p/m_e)\Gamma(\theta)$, where $g(p)=(p-2)/(p-1)$ for $p>2$. The cooling Lorentz factor is defined as $\gamma_c(\theta, r,t) = \frac{6\pi m_e c}{(1+Y)\,\sigma_T\,B'(\theta,r)^2\,\Gamma(\theta)\,t/(1+z)}$, where $Y$ is the Compton parameter representing the SSC-to-synchrotron luminosity ratio, $\sigma_T$ is the Thomson cross section, $t$ is the observer time.

At relatively low frequencies, the Compton parameter $Y$ is assumed to remain in the Thomson regime \citep{mondal2023probing, mondal2025follow}. For a given values of  $\gamma_m$ and $\gamma_c$, the synchrotron spectrum is characterized by the standard break frequencies--- the injection frequency $\nu_{m}(\gamma_{m})$, the cooling frequency $\nu_{c}(\gamma_{c})$, and the maximum cutoff frequency $\nu_{max}(r)$ \citep{mondal2023probing, mondal2025follow}.

We define the synchrotron peak flux density for each jet segment, which is modified by the presence of the wind density profile 
$n(r)$ as, 

\begin{equation}
    F_{max}(R) = \frac{(1+z)}{d_{L}^{2}}\int_{0}^{R}n(r) 4 \pi r^{2} P_{\nu,max}(r) dr
\end{equation}
where, $P_{\nu,max}$ is the synchrotron peak power per unit frequency. Thus, the synchrotron peak flux density for each jetted segment is obtained as,

\begin{equation}\label{eqn_ssc_peak}
 F_{\nu,max}^{(i,k)}(R)= n_{k} \times R \times P_{\nu,max}^{(i,k)}(R) \times \frac{\Omega_{i,k}}{\Omega_{e,i,k}} \times \frac{(1+z)}{d_{L}^{2}}.   
\end{equation}

Where $R$ is the shock front radius associated with the segment's deceleration radius $r_{dec}$. The ratio $\frac{\Omega_{i,k}}{\Omega_{e,i,k}}$ accounts for the fraction of the jet solid angle contributing to the emission observed from that segment following \citet{lamb2017electromagnetic}.

Hence, the observed (off-axis) synchrotron flux for each jet segment associated with its on-axis flux is calculated as \citep{lamb2017electromagnetic},

\begin{equation}
\begin{aligned}
F_{\rm syn}^{(i,k)}\!\left(t_{\rm obs},\alpha_{\rm inc}^{i,k},R\right)
&= \delta_{\rm dop}^{3}\!\left(r,\alpha_{\rm inc}^{i,k}\right)\,
\cos\!\alpha_{\rm inc}^{i,k}\,
F_{\nu,\max}^{(i,k)}(R)
\\
&\quad\times f_{\nu_{syn}}.
\end{aligned}
\end{equation}

Here, $f_{\nu_{syn}}$ represents normalised synchrotron spectra, which have two separate spectral profiles for slow and fast cooling regimes \citep{Sari_1998, 2018pgrb.book.....Z}. For each surface element $(\theta_{i},\phi_{k})$, the observed specific flux is modified by the Doppler factor $\delta_{dop}(r,\alpha_{inc}^{i,k}) =\frac{1 - \beta(r)}{1 - \beta(r) \cos \alpha_{i,k}}$, and the frequency related to the on-axis observed is also modified as $\nu_{off}^{(i,k)}= \delta_{dop}(r,\alpha_{inc}^{(i,k)}) \nu_{on}$.

\subsection{Estimating SSC Flux}

Following \citet{mondal2025follow}, the observed SSC flux for each jet segment $(i,k)$ is estimated similarly as ---

\begin{equation} \label{eqn_9}
\begin{aligned}
     F_{SSC}^{(i,k)}(r) &= F_{\nu,max}^{(i,k)}(R) \times (n_{k}R\sigma_{T}x_{0}) \times f_{\nu_{SSC}}.
\end{aligned}
\end{equation}

Here $F_{\nu,max}^{(i,k)}(R)$ represents maximum synchrotron peak flux along radius R for each of the jet segments, $n_{k}$ is the wind density normalisation constant, and $\sigma_{T}$ is the Thomson scattering cross-section. We computed the on-axis SSC flux profile $f_{\nu_{SSC}}$ in the slow and fast cooling regimes following \citet{sari2001synchrotron} and \citet{zhang2018physics}. Additionally, we adopted a value of constant parameter $x_{0}$ as $0.5$, as the fractional scattering efficiency in the Thomson regime~\citep{sari2001synchrotron}. 
 
At very high energies, SSC scattering enters the Klein–Nishina (KN) regime, which leads to suppression of the high-energy emission and modifies the spectral break structure \citep{nakar2009klein}. We incorporate KN corrections into our structured jet wind model following the numerical approach presented in \citet{mondal2025follow}, which accounts for the transition from the Thomson to the KN regime, introducing additional critical Lorentz factors and spectral breaks in the SSC component. In the wind-driven medium, due to the modified radial dependencies of the minimum and cooling electron Lorentz factors $\gamma_{m}$ and $\gamma_{c}$, the critical Lorentz factors themselves are modified as $\hat{\gamma}_{m}$ and $\hat{\gamma}_{c}$. Correspondingly, the Compton parameter $Y$ in the KN regime for slow cooling, ($Y({\gamma_{c}})$), and fast cooling regime ($Y({\gamma_{m}})$) are modified following the prescription in \citet{mondal2025follow}. Additionally, we include attenuation of VHE gamma-ray flux due to extragalactic background light (EBL) absorption using the Domínguez EBL model~\citep{dominguez2011extragalactic}. This correction is crucial for accurately modeling the observed VHE flux from GRB afterglows in a wind-driven medium.

\section{Impact of Gaussian structured jet and afterglow parameters on characteristics of TeV light curve}\label{sec:param_charc}

\begin{figure}[ht]
    \centering
    \includegraphics[width=0.9\linewidth]{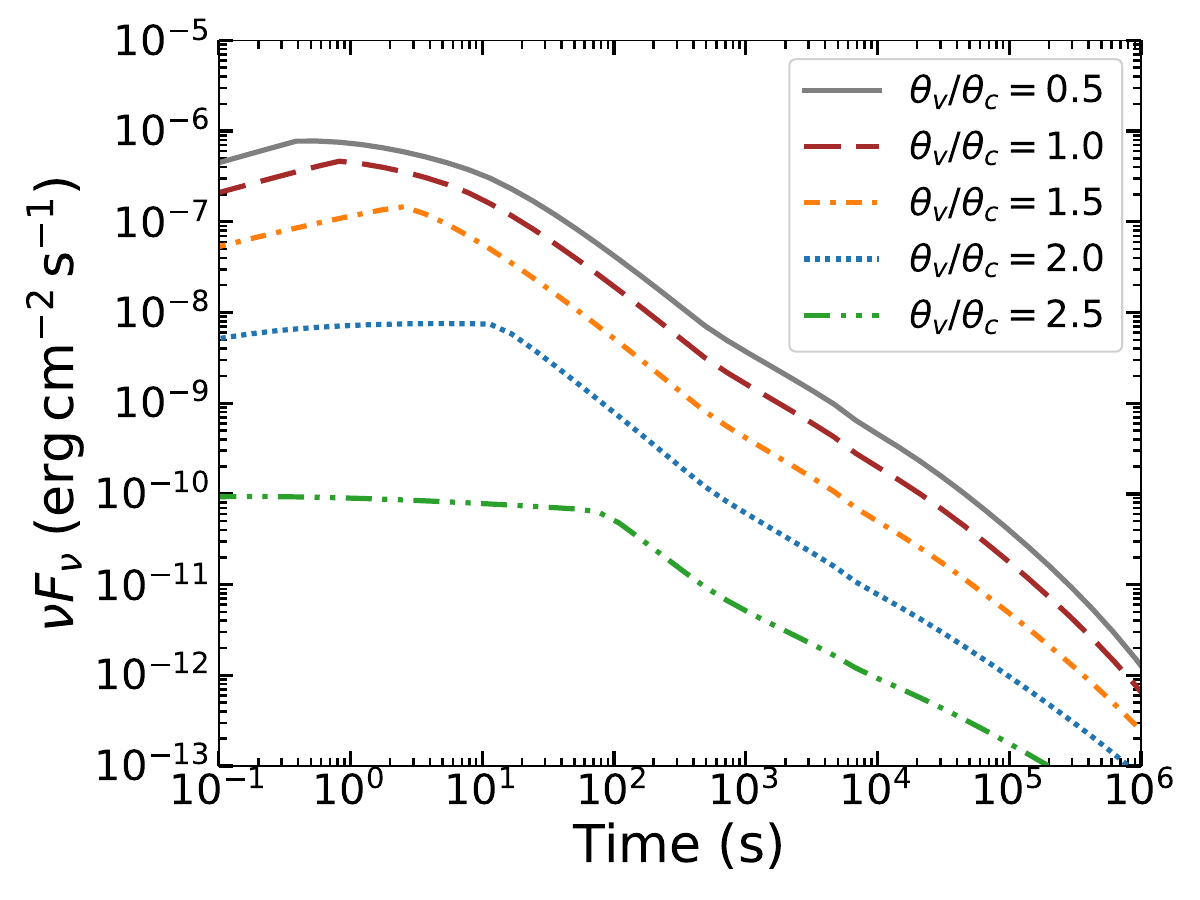}
    \caption{Time evolution of the SSC afterglow flux at $z=0.151$ is plotted for different jet geometry ($\theta_{v}/\theta_{c}$). We have kept $\theta_{c} $ fixed at $6^{\circ}$ and the curves correspond to varying viewing angles $\theta_v=\{3^\circ,\,6^\circ,\,9^\circ,\,12^\circ,\,15^\circ\}$, corresponds to typical on-axis to off-axis jet scenario. The afterglow parameters are fixed to $E_k=5.5\times10^{52}\,\mathrm{erg}$, $\eta_0=440$, $A_{\star}=0.56$, $\epsilon_e=0.1$, $\epsilon_B=10^{-3}$ and $p=2.5$. All of these SSC fluxes include the EBL correction.}
    \label{fig:thv_thc_ratio}
\end{figure}

\begin{figure*}[ht]
  \centering
 \subfloat[\label{fig:Ek_Astar_gam0_on_off_LC1}]{\includegraphics[width=0.32\linewidth]{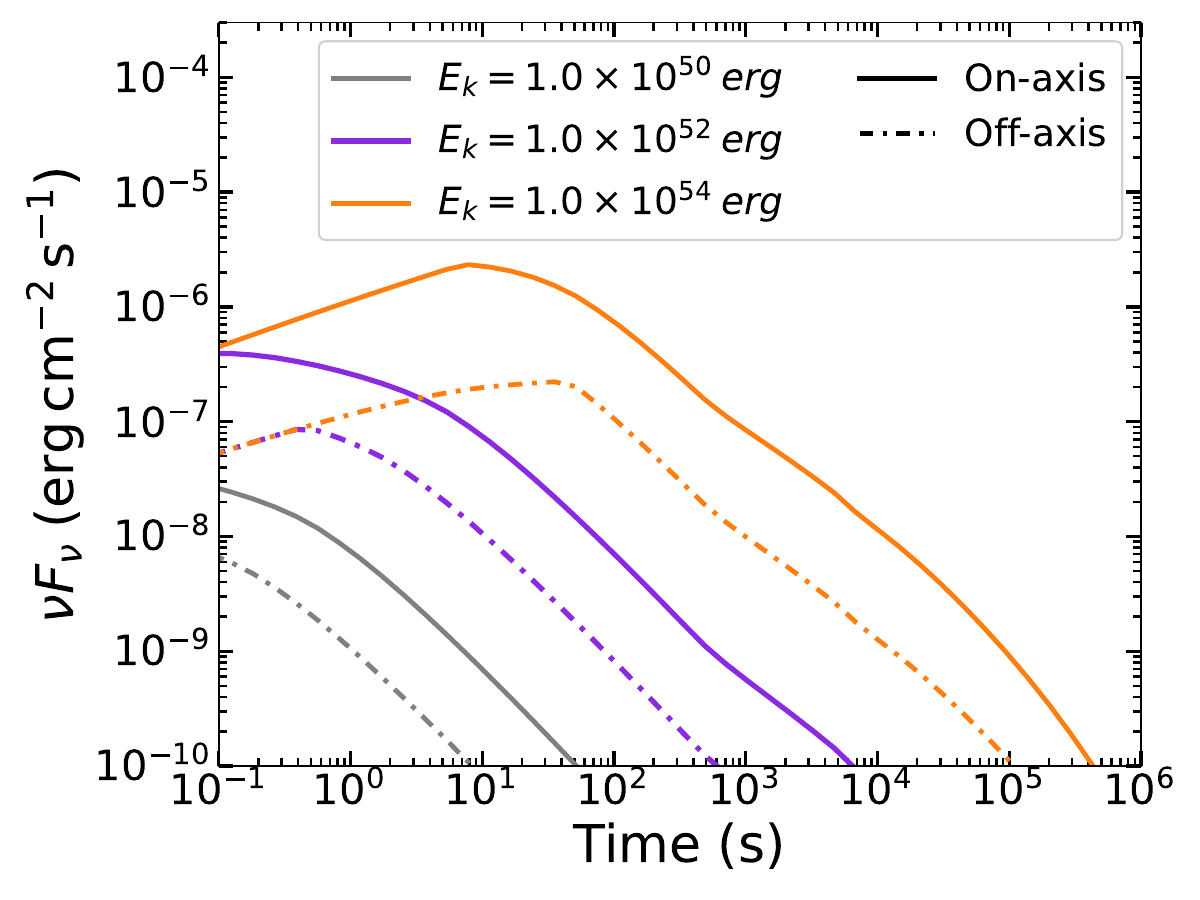}}
 \subfloat[\label{fig:Ek_Astar_gam0_on_off_LC2}]{\includegraphics[width=0.32\linewidth]{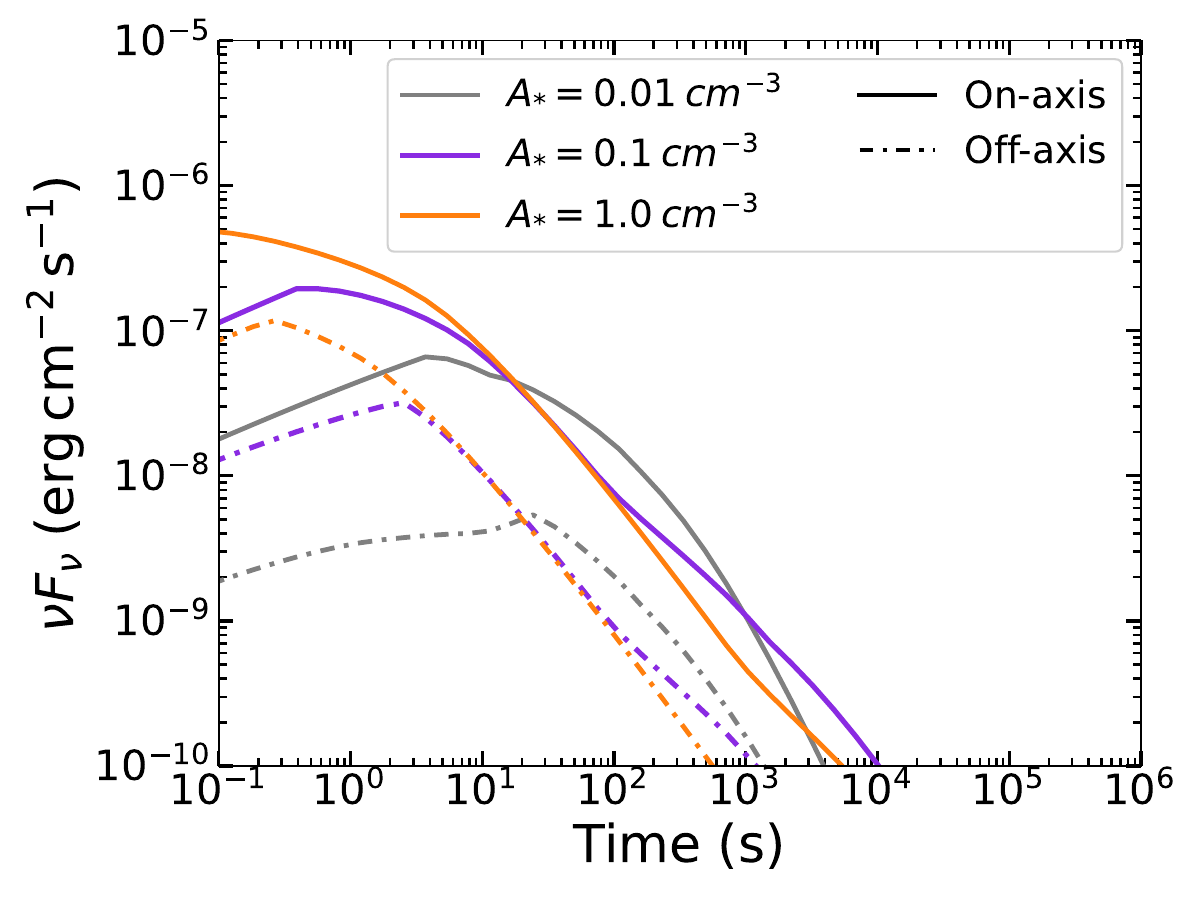}}
 \subfloat[\label{fig:Ek_Astar_gam0_on_off_LC3}]{\includegraphics[width=0.32\linewidth]{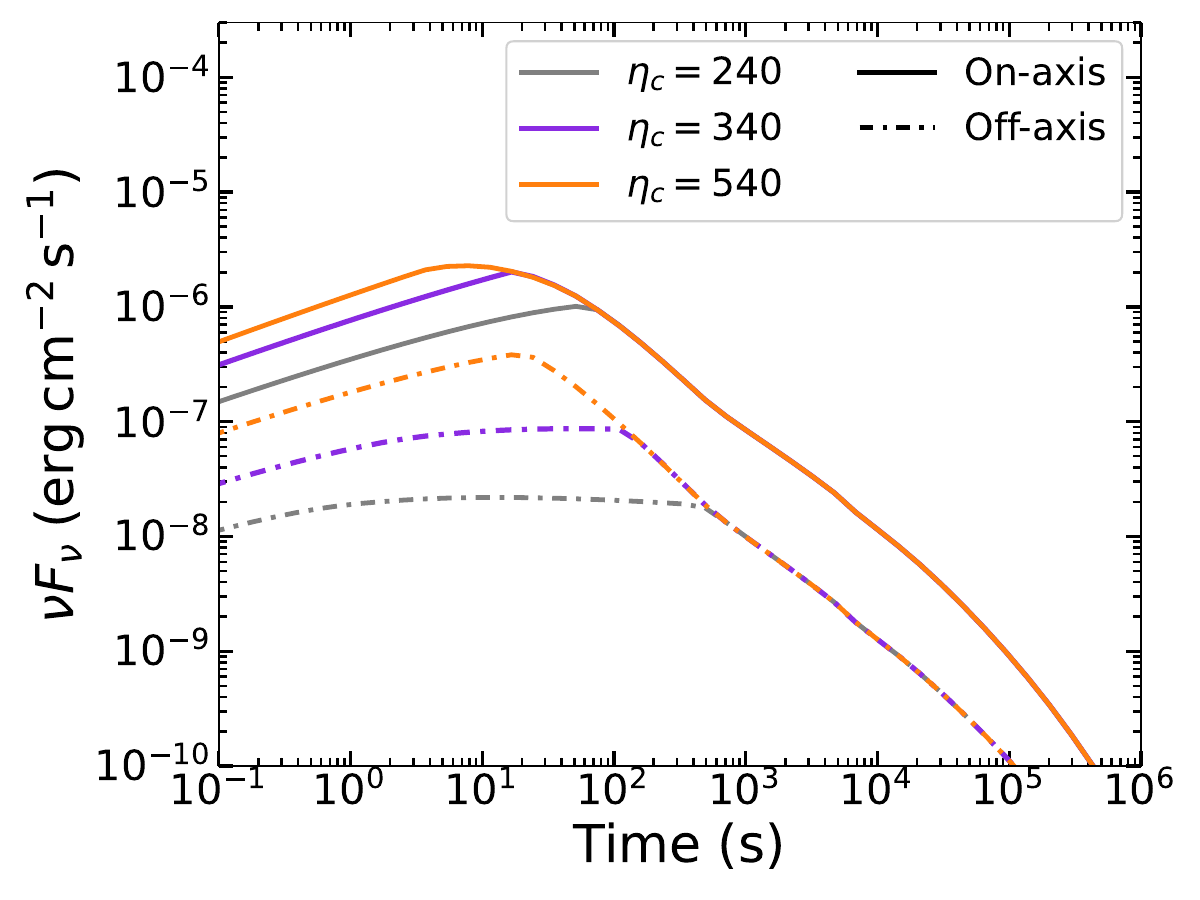}}
  \caption{We plot SSC afterglow flux at $z=0.151$ with varying parameters as in (a) $E_k = \{10^{50}, 10^{52},10^{54}\}\,\mathrm{erg}$, (b) $A_{\star}=\{0.01,\,0.1,\,1.0\}$ and (c) $\eta_c=\{240,\,340,\,540\}$. Otherwise parameters are fixed to $E_k=10^{52}\,\mathrm{erg}$, $\eta_0=440$, $A_{\star}=0.56$, $\epsilon_e=0.1$, $\epsilon_B=10^{-3}$ and $p=2.5$. For the on-axis case (solid curves)  $\theta_v=3^\circ$ and for the off-axis case (dashed-dot curves) $\theta_v=9^\circ$, whereas $\theta_{c} $ is kept fixed at $6^{\circ}$. The light-curve plot depicts the SSC flux with EBL correction.}
  \label{fig:Ek_Astar_gam0_on_off_LC}
\end{figure*}

\begin{figure}[ht]
    \centering
    \includegraphics[width=0.9\linewidth]{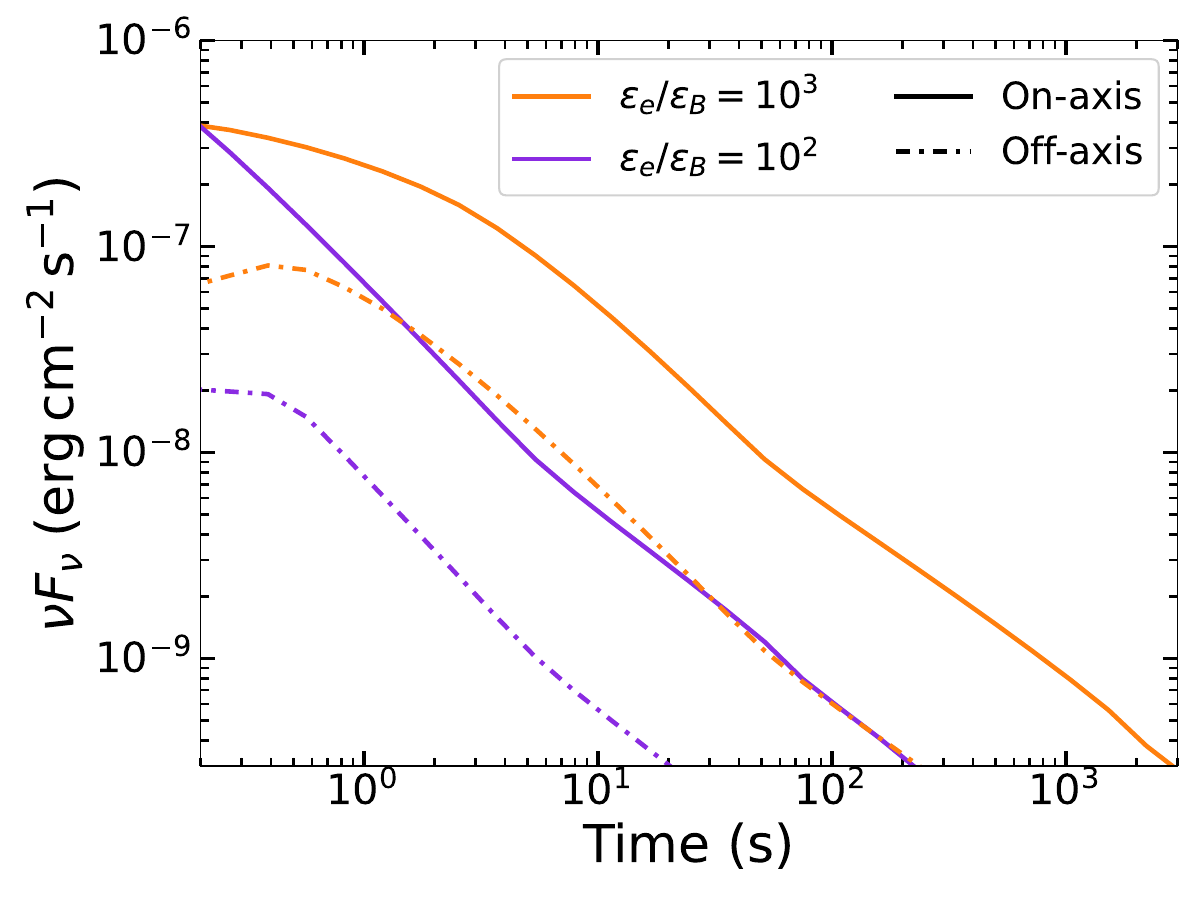}
    \caption{SSC afterglow light curves at $z=0.151$ for fixed $\epsilon_B=10^{-3}$ and varying $\epsilon_e= \{0.1,\;0.01\}$. Other afterglow parameters are fixed to $E_k=10^{52}\,\mathrm{erg}$, $\eta_0=440$, $\epsilon_e=0.1$, $\epsilon_B=10^{-3}$, and $p=2.5$. We show on-axis, solid curves ($\theta_v=3^\circ$) and off-axis, dashed-dot curves ($\theta_v=9^\circ$) cases kepeping fixed $\theta_c$ at $6^\circ$. Fluxes include attenuation by EBL correction.}
    \label{fig:epse_epsB_ratio}
\end{figure}

In this section, we analyse the characteristics of TeV afterglows evolving in a stellar-wind medium, as detailed in Section~\ref{sec:model}. Our wind afterglow model is primarily governed by the jet structure parameters $\theta_{c}$, and $\theta_{v}$, along with the afterglow parameters $E_{k}$, $A_{\star}$, $\epsilon_{e}$ and $\epsilon_{B}$. In the following sections, we discuss how variations in each of these parameters shape the modeled light curve, while isolating their individual and combined influences on the TeV evolution.

\subsection{Dependence on jet structure parameters}
\label{sec:depend_thv_thc}

Unlike a uniform interstellar medium (ISM), the wind density profile modifies both the deceleration dynamics and the beaming evolution of the blast wave. To isolate the effect of jet structure on the TeV afterglow light curve, we hold all parameters fixed and systematically vary only the ratio $\theta_v/\theta_c$. The resulting light curves, shown in Figure~\ref{fig:thv_thc_ratio}, reveal the influence of angular viewing geometry on the timing and intensity of the TeV emission peak.

For a structured jet, an observer at viewing angle $\theta_v$ sees different portions of the angular energy profile, as the outflow decelerates and the lateral beaming widens. The relativistic Doppler factor $\delta_{dop}(r,\alpha_{inc}^{i,k})$ governs the observed peak flux and peak time. Consequently, the ratio $\theta_v/\theta_c$ spontaneously captures the light-curve features. Core-aligned or on-axis views ($\theta_v<\theta_c$) produce earlier, brighter peaks, whereas off-core or off-axis views ($\theta_v>\theta_c$) create dimmer light curves that peak at later epochs as the beaming cone widens to confine the line of sight.

As we see in section~\ref{sec:model}, this angular dependence in a wind medium is essential because $r_{\rm dec}$ scales differently than in the ISM, shifting peak times. Moreover, at TeV energies the emission is typically SSC-dominated, and both the beaming and the seed synchrotron field are sensitive to the jet’s angular structure. As a result, CTA detectability can vary by orders of magnitude depending on whether the observer is core-aligned or off-axis in a wind environment, making $\theta_v/\theta_c$ a key predictor of observational outcome.

Figure~\ref{fig:thv_thc_ratio} displays TeV light curves for five cases with $\theta_v/\theta_c=\{0.5,\,1.0,\,1.5,\,2.0,\,2.5\}$ while all other parameters are held fixed at $E_k = 5.5\times10^{52}\:erg$; $A_{\star}=0.56$, $\epsilon_{e}=0.1$, $\epsilon_{B}=0.001$, $\eta_{c}=440$, $p=2.45$. This study is conducted at redshift $z=0.151$. Two robust trends emerge here as discussed above. When the viewing angle is within the jet core ($\theta_v<\theta_c$), strong Doppler boosting leads to an early and brighter peak in the TeV light curve; this corresponds to core-aligned geometries (e.g. $\theta_v/\theta_c=0.5$, solid grey curve in Figure~\ref{fig:thv_thc_ratio}), where the observer is closely aligned with the jet axis. In this regime, the combination of intense Doppler boosting and higher jet core kinetic energy per unit solid angle produces an early peak near the onset of deceleration, with peak time $t_{\rm pk}\!\sim\!t_{\rm dec}(\theta\!\approx\!0) \propto \mathcal{E}_c/(A_{\star}\,\Gamma_{c}^4)$. As the viewing angle increases and the ratio $\theta_v/\theta_c$ increases from $1.0$ to $2.5$ (dashed brown to dash-dot green), the light curve peaks shift to later times and the observed flux declines by several orders of magnitude. 

In figure~\ref{fig:thv_thc_ratio}, when the viewing angle lies outside the jet core ($\theta_v\!>\!\theta_c$), the effective observer angle $\theta_{v}$ becomes larger, lowering the Doppler factor $\delta_{\rm dop}$. This leads to enhanced Doppler de-boosting, which both delays the time at which the jet emission becomes visible to the observer (approximately when $\Gamma\!\sim\!1/\theta_v$) and reduces the observed SSC emissivity. Furthermore, both the kinetic energy per unit solid angle, $\mathcal{E}(\theta_v)$ and $\Gamma_0(\theta_v) \beta_{0}(\theta_{v})$ decrease exponentially as  $\theta_v/\theta_c$ increases. This exponential decline further reduces the radiating electron population and the strength of comoving magnetic fields that govern SSC emission. Thus with increasing $\theta_v/\theta_c$, in a wind-driven environment, the peak time becomes $t_{\rm pk}\!\propto\!\mathcal{E}(\theta_v)/[A_{\star}\,\Gamma_0(\theta_v)^4]$, amplifying the shift of the light curve peak to later times as off-axis viewing becomes more pronounced. This reflects the combined effect of reduced Doppler boosting and decreasing on-axis energy, resulting in fainter and more delayed emission when the observer is progressively misaligned with respect to the jet core.

At late epochs, $t \gg t_{\rm pk}(\theta_v)$, the blast wave becomes only mildly relativistic, and the beaming dependence on $\theta_v$ weakens. Consequently, the TeV light curves for different $\theta_v/\theta_c$ converge toward similar temporal decay slopes. These slopes are set primarily by the adiabatic evolution inherent to the wind density profile and by the overall energy of the jet. This convergence occurs more slowly than in a uniform ISM environment because, within a wind medium where the swept-up mass grows linearly, $m(r)\propto r$, the bulk Lorentz factor declines more gradually. Therefore, angular variations in jet energy $\mathcal{E}(\theta)$ and jet Lorentz factor $\Gamma_0(\theta)$ persist for a longer period before being smoothed out.

\subsection{Dependence on afterglow parameters}
\label{sec:Ek_Astar}

Figure~\ref{fig:Ek_Astar_gam0_on_off_LC1} and \ref{fig:Ek_Astar_gam0_on_off_LC2} represents how the TeV light-curve peak time and peak flux depends on variations of the total jet kinetic energy $E_k$ and the wind density parameter $A_{\star}$, for on-axis ($\theta_v<\theta_c$) and off-axis ($\theta_v>\theta_c$) views. In all panels, the other parameters are fixed, except for those being varied. This dependency can be understood in terms of the wind deceleration epoch and Doppler beaming. As, we discussed earlier (section~\ref{sec:depend_thv_thc}) that, for a wind ($n\propto r^{-2}$), the deceleration time at polar angle $\theta$ scales as $t_{\rm dec}(\theta)\ \propto\ \frac{\mathcal{E}(\theta)}{A_{\star}\,\Gamma_0(\theta)^4}$, and SSC-dominated flux at given epoch is strongly govened by beaming $\propto \delta_{dop}^{3}$.

Increasing $E_k$ raises $\mathcal{E}(\theta)$ at all polar angles for all jet segments, which increases the radiating nonthermal electrons and thus boosts the seed photon field by increasing SSC flux. The peak time of the on-axis ($\theta_v<\theta_c$) observer is almost equivalent to the deceleration time epoch, which is directly proportional to $E_{k}$. Consequently, the peak shifts to later times at higher flux. Thus, Figure~\ref{fig:Ek_Astar_gam0_on_off_LC1} depicts a monotonic rise in peak flux with increasing $E_{k}$ together with a rightward shift of the peak turnover. In the off-axis ($\theta_v>\theta_c$) scenario (Figure~\ref{fig:Ek_Astar_gam0_on_off_LC2}), the same intrinsic trend is present but partially dominated by Doppler deboosting. Hence, the light curve peak to an off-axis observer is delayed by $\Gamma(\theta_{v})\!\sim\!1/\theta_v$, so peaks are systematically at later and dimmer than their on-axis counterparts for the same $E_k$. In short, a larger $E_k$ makes the afterglow intrinsically brighter and pushes its peak to later times, while the viewing angle governs the Doppler beaming.

A larger $A_{\star}$ increases the external density at a given radius in the wind medium, which further increases the number of swept-up electrons and strengthens the comoving magnetic and photon energy densities, enhancing the SSC flux. However, at the same time, increasing $A_{\star}$, advances the deceleration epoch because $t_{\rm dec}\!\propto\!A_{\star}^{-1}$, which leads to an earlier and brighter SSC peak. This effect is most visible for $\theta_v<\theta_c$ where Doppler boosting is maximal. 
In the off-axis scenario, the peak remains delayed relative to the on-axis case, which introduces an additional geometric timescale due to enhanced Doppler deboosting, keeping the peak flux below that of the on-axis case even when $A_{\star}$ is high. 

In Figure ~\ref{fig:Ek_Astar_gam0_on_off_LC3} we observe that, for On-axis scenario ($\theta_v<\theta_c$), larger $\eta_c$ increases the early-time beaming and advances the peak because $t_{\rm pk}\!\propto\!\Gamma_c^{-4}$.  This is seen as the orange ($\eta_c=540$) curve, which peaks slightly earlier and at a higher flux than the purple ($\eta_c=340$) and grey ($\eta_c=240$) curves, respectively. 
After the peak turnover, the curves converge as the flow becomes only mildly relativistic and beaming differences diminish, as discussed earlier. However, for the off-axis ($\theta_v>\theta_c$) scenario, with the same value of $\Gamma_c$, peak flux lowers compared to the on-axis case, and observed peak shifts to later time. Accordingly, the orange dashed-dot curve is brighter but peaks earlier than the lower $\Gamma_c$ cases in the off-axis scenario. 

Figure~\ref{fig:epse_epsB_ratio} shows TeV light curves for two distinct ratios of microphysical parameters $\epsilon_e/\epsilon_B$. With $\epsilon_B$ held fixed, increasing the ratio $\epsilon_e/\epsilon_B$ means we vary $\epsilon_{e}$ to its higher and lower values i.e., $\epsilon_e=\{10^{-2},10^{-1}\}$. All solid curves represent on-axis ($\theta_v<\theta_c$) and dash–dot depicts off-axis ($\theta_v>\theta_c$) scenario. The raising $\epsilon_e/\epsilon_B$ primarily boosts the SSC peak flux because of the increased Compton Y parameter~\citep{mondal2023probing, mondal2025follow}, and $\gamma_m\!\propto\!\epsilon_e$ (see section~\ref{subsec:wind_effects}) elevates both the synchrotron seed photon field and the SSC flux. Consequently, the higher–ratio $\epsilon_e/\epsilon_B=10^3$, orange LCs are brighter than the lower–ratio $\epsilon_e/\epsilon_B=10^2$, purple LCs. Only the peak time changes slightly, since varying $\epsilon_e$ keeps the deceleration epoch unchanged. Thus, the geometry of the jet, rather than the microphysical parameters, primarily determines the variations in peak time and peak flux.

\section{Detectability of TeV light curves in wind-driven medium with CTA sensitivity}\label{sec:detect_TeV_events}

\begin{figure*}
\centering
\includegraphics[width=0.3\linewidth]{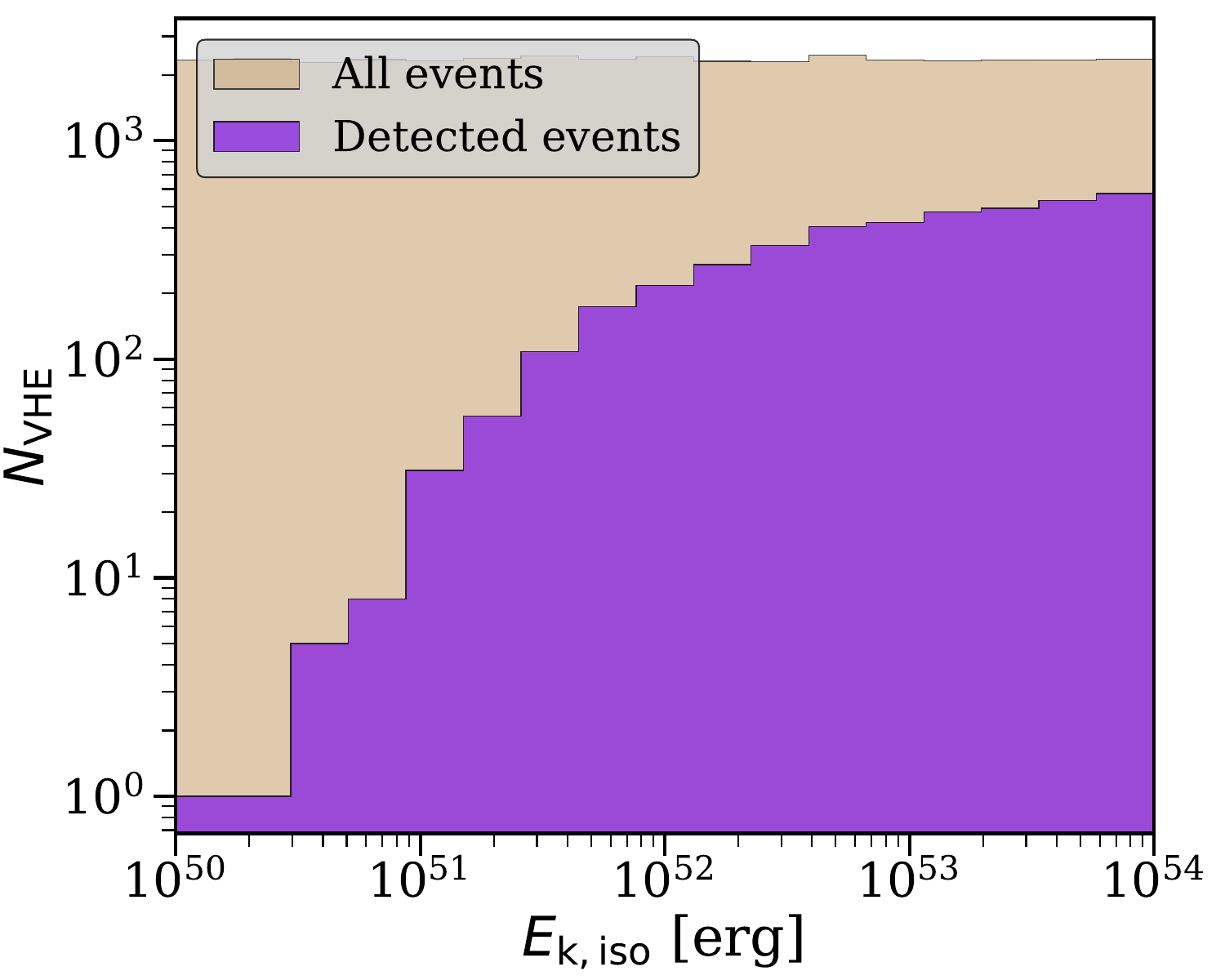} 
\includegraphics[width=0.3\linewidth]{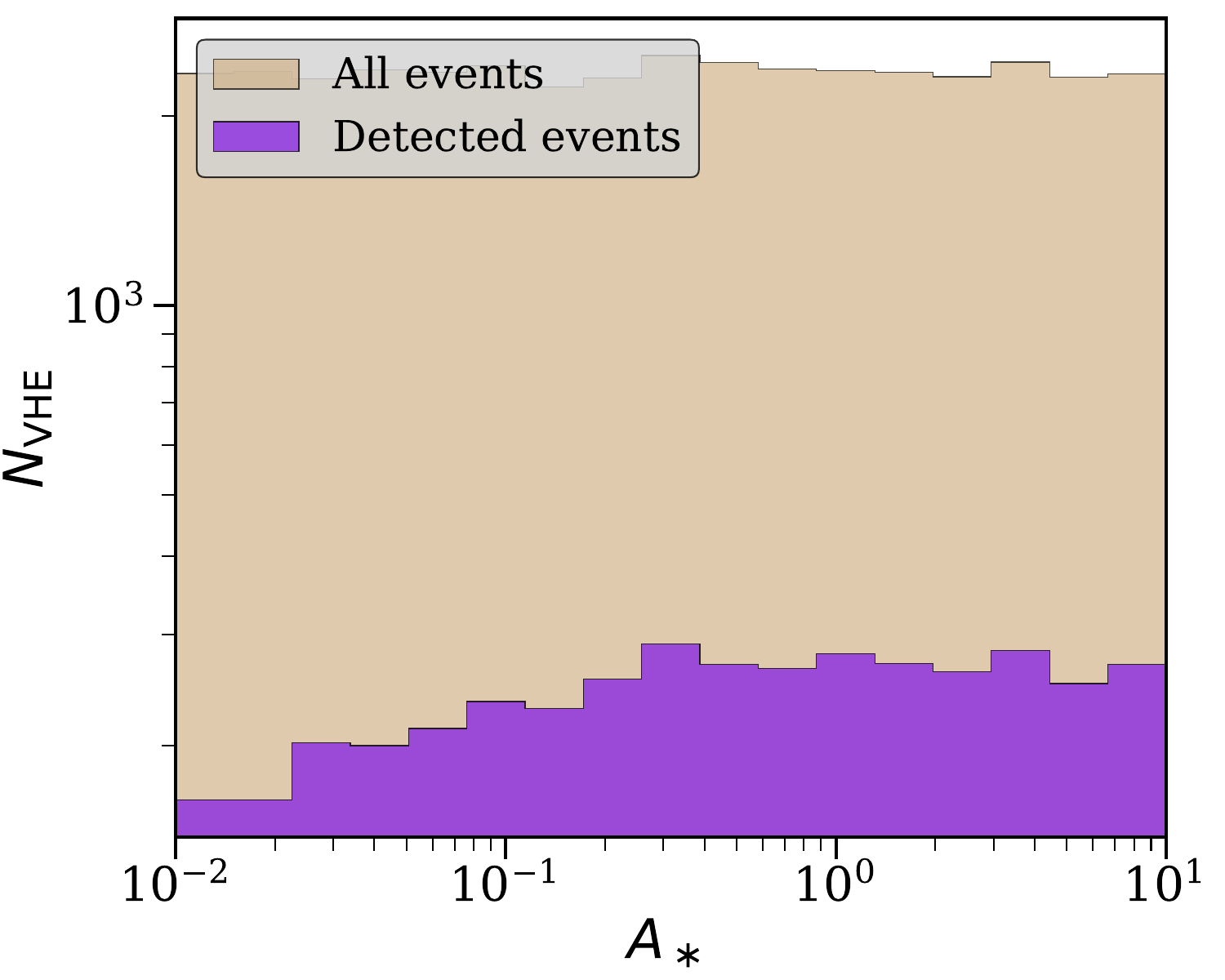}
\includegraphics[width=0.3\linewidth]{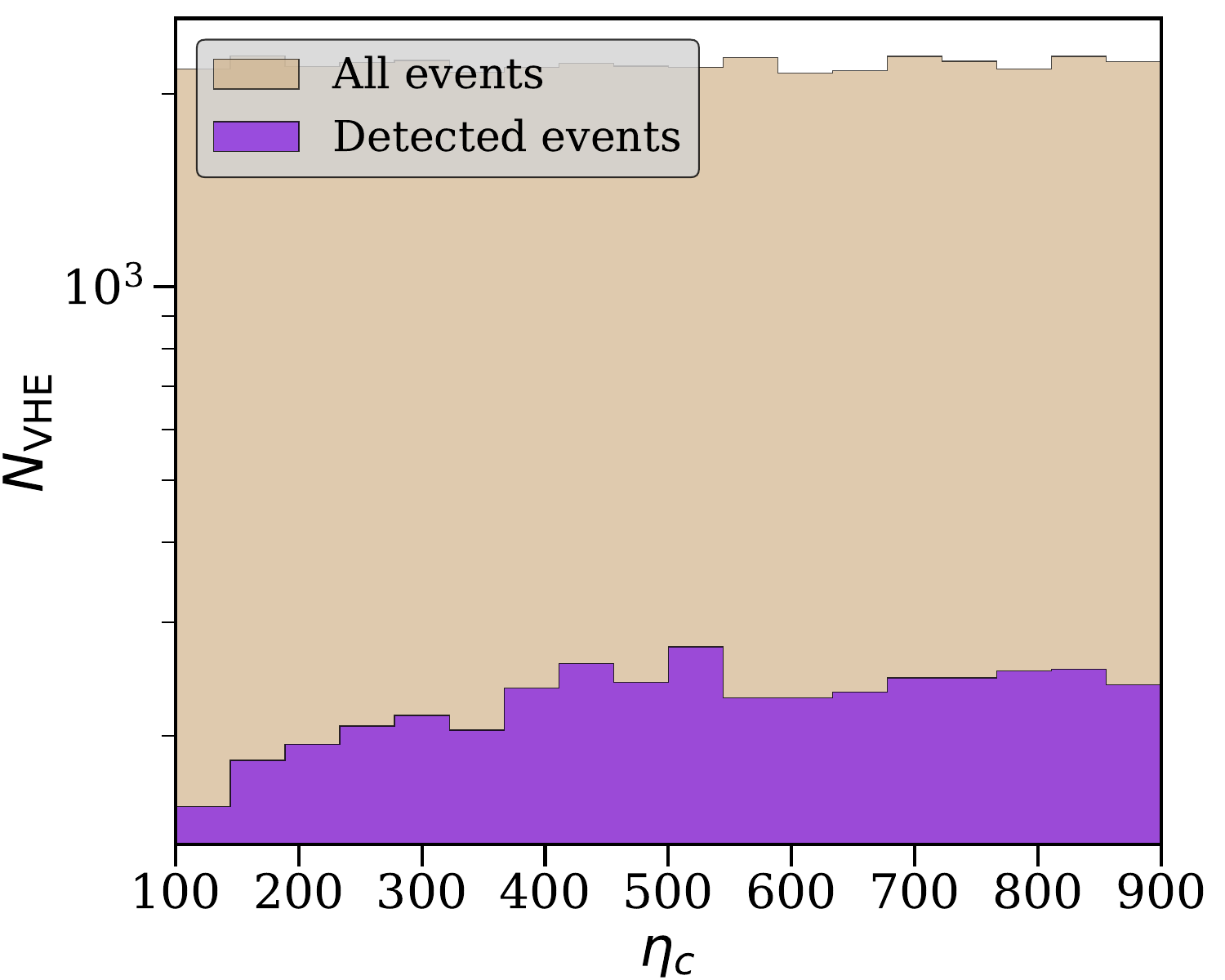}
\caption{Histograms showing the distribution of all CTA-detected events compared to all simulated samples for parameter $E_k$, $A_{\star}$ and $\eta_c$. These plots depict that higher values of $E_{k}$ and $A_{\star}$ are more probable for CTA detections. Whereas moderately high values of $\eta_c$ favour CTA detectability.}
\label{fig:Hist_Ek_A_etac_dL}
\end{figure*}

\begin{figure}
\centering
\includegraphics[width=0.45\linewidth]{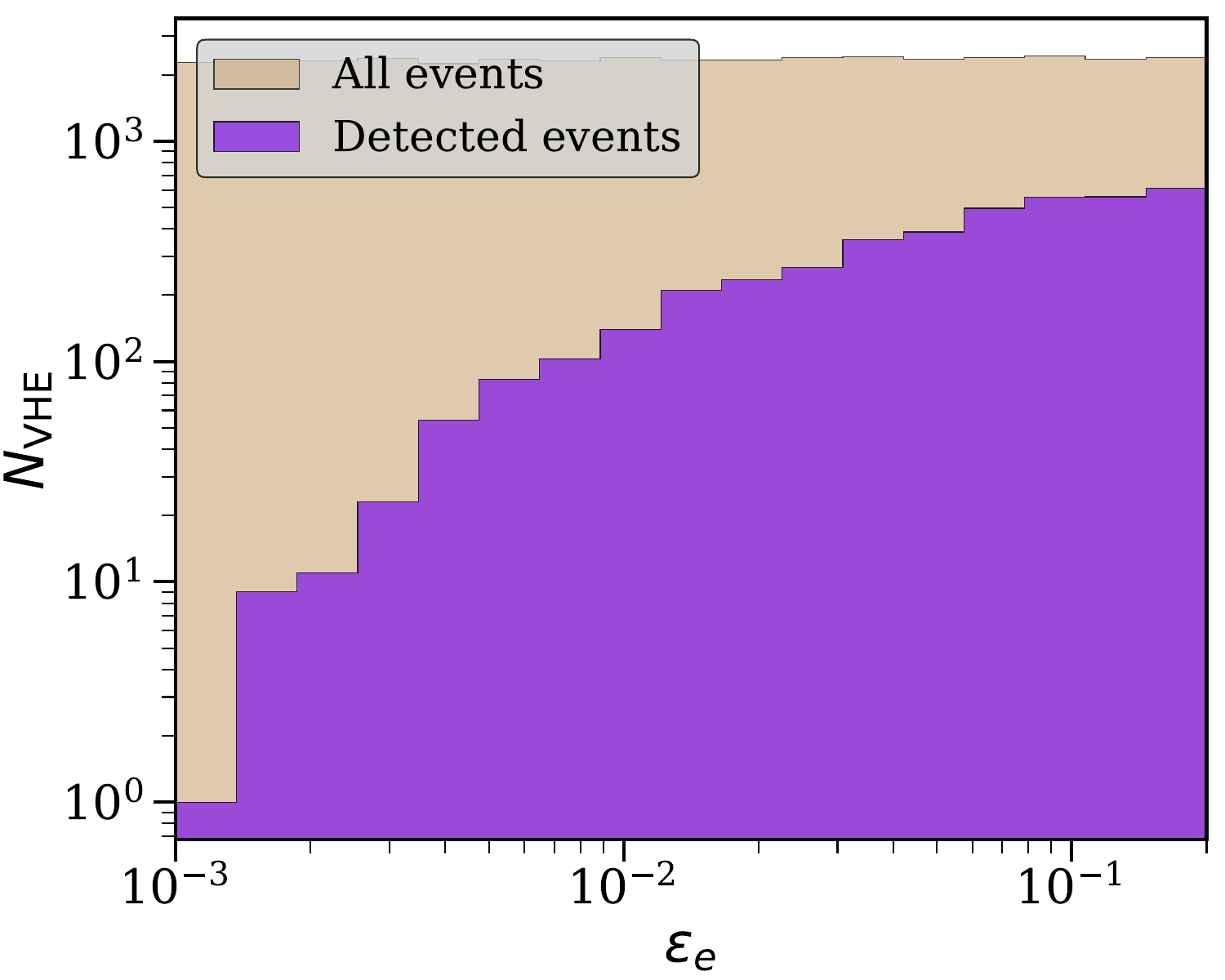} 
\includegraphics[width=0.45\linewidth]{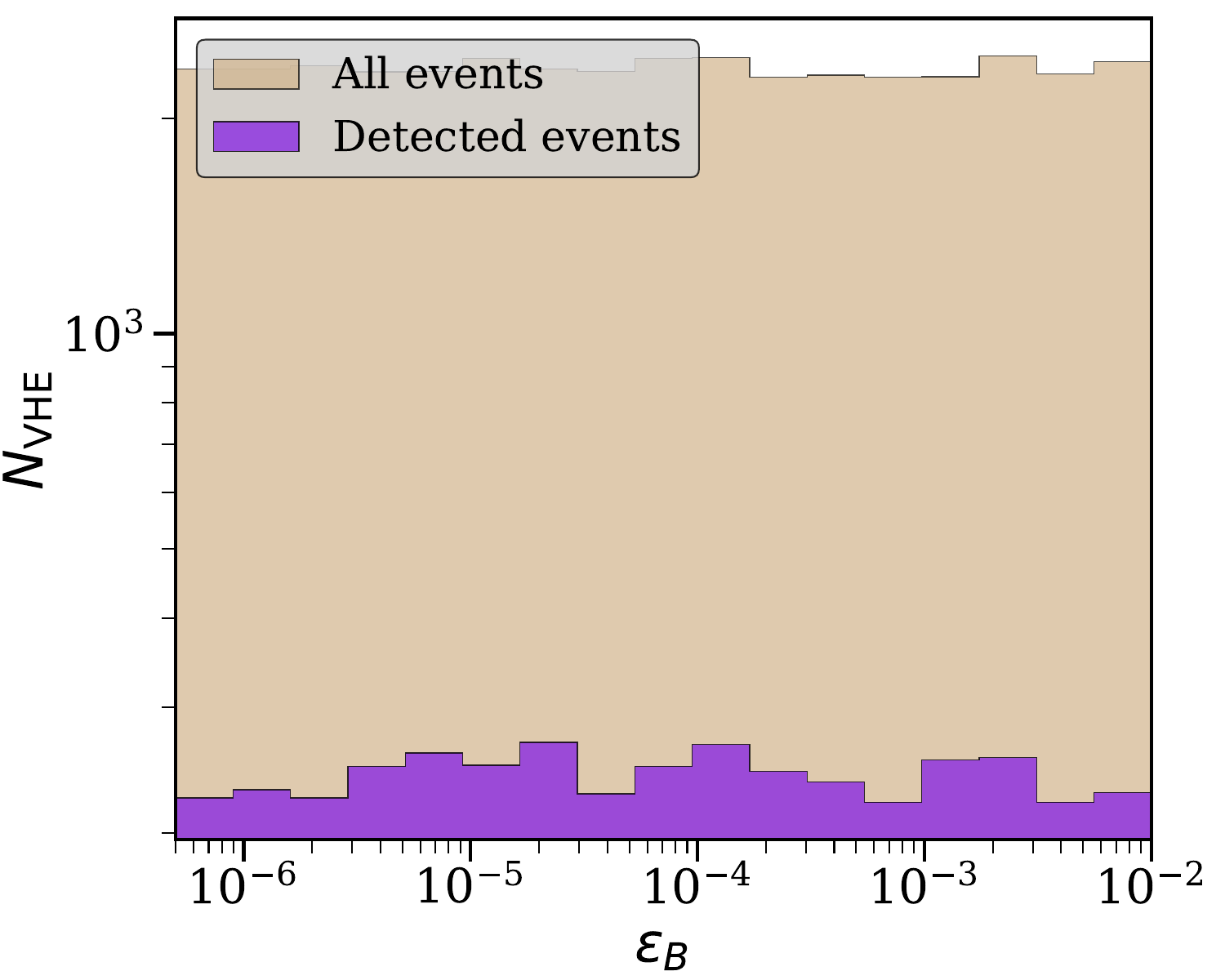}
\caption{Histograms illustrate all CTA detected events are more probable at higher values of $\epsilon_{e}$ (left), whereas the nearly uniform distribution in $\epsilon_B$(right) indicates that this parameter has a weaker influence on CTA detectability.}
\label{fig:Hist_epse_epsB}
\end{figure}

Building on the Gaussian structured jet model in a wind medium presented in Section~\ref{sec:model}, we now investigate the detectability of simulated GRB afterglow light curves using the differential flux sensitivity of the CTAO Northern array\footnote{\url{https://www.ctao.org/for-scientists/performance/}} at 250~GeV, for integration times ranging from a few seconds to a few hours.
 To explore the detectability of TeV afterglow light curves in a stellar wind medium, with the CTA, we simulated $4\times10^{4}$ events over a broad prior range of seven afterglow parameters. Similar to \citet{mondal2025follow}, an event is considered to be detected if the light curve crosses the CTA's detection threshold at any point in time. The luminosity distance of all those events is fixed at $z=0.5$ ($d_{L}\approx 2863.1$ Mpc). All SSC light curves include EBL attenuation. 

Our afterglow parameter space comprises the following parameters --- jet kinetic energy $E_{k}$, wind number density $A^{*}$, fractional kinetic energy transferred to electron and magnetic field $\epsilon_{e}$ and $\epsilon_{B}$, jet structure parameter $\theta_{c}$, jet viewing angle $\theta_{v}$, and initial velocity $\eta_{c}$. For each simulated event, we choose a random value for every parameter from the specified prior distributions. We adopt log-uniform priors for $E_k$, $A_{\star}$, $\epsilon_e$, and $\epsilon_B$, with ranges
$E_k \in [10^{50},10^{54}]~\mathrm{erg}$, $A_{\star} \in [10^{-3},1]~\mathrm{cm}^{-3}$, $\epsilon_e \in [10^{-3},0.2]$, and $\epsilon_B \in [5\times10^{-7},10^{-2}]$. The angular parameters are taken to be uniformly distributed, with $\theta_c$ in the range $1^{\circ}$–$13^{\circ}$ and $\theta_v$ in the range $0^{\circ}$–$20^{\circ}$, while the jet opening angle is fixed at $\theta_j = 15^{\circ}$. The initial Lorentz factor $\eta_c$ is drawn from a uniform prior between 100 and 900.

\begin{figure*}
\centering
\includegraphics[width=0.32\linewidth]{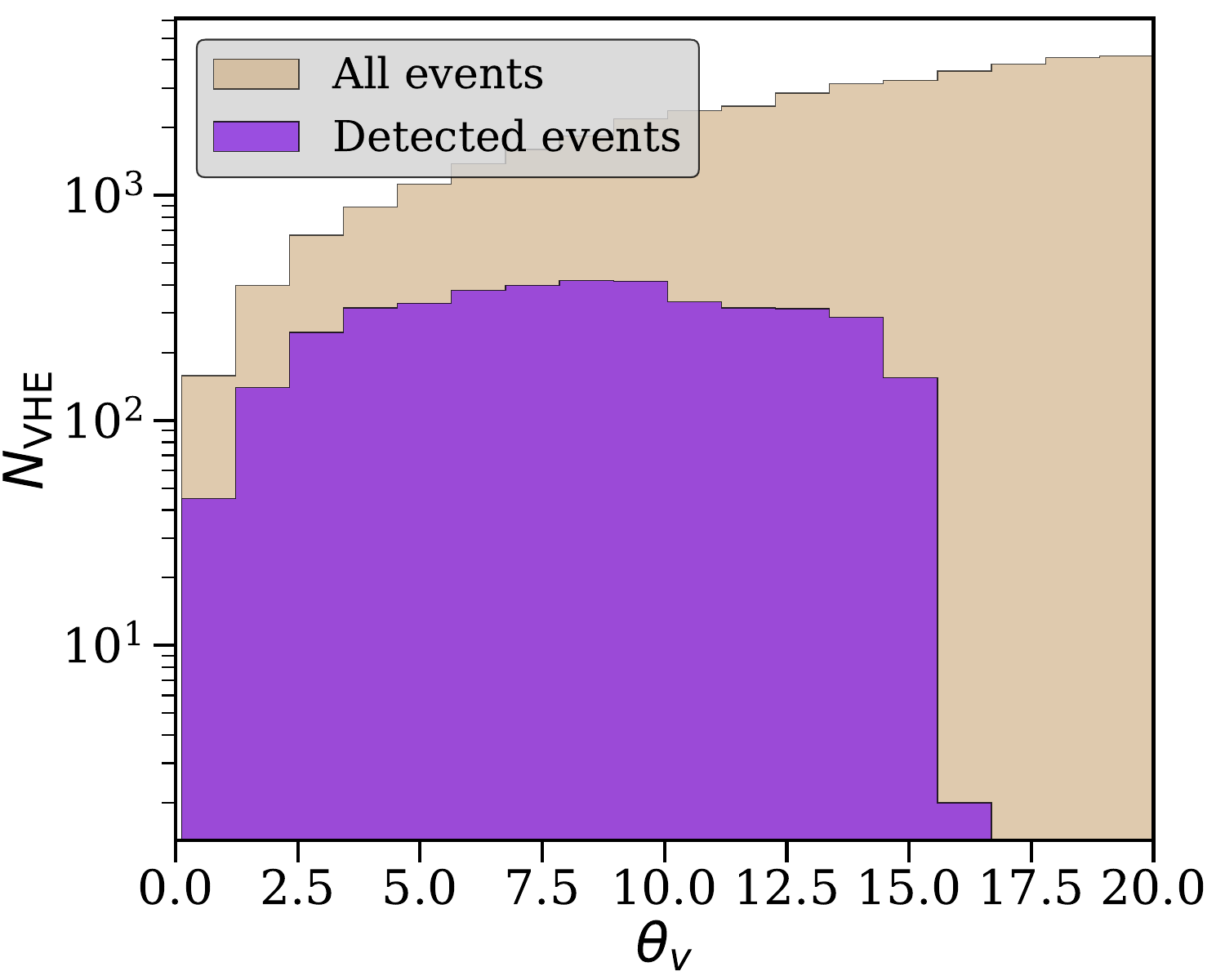} 
\includegraphics[width=0.32\linewidth]{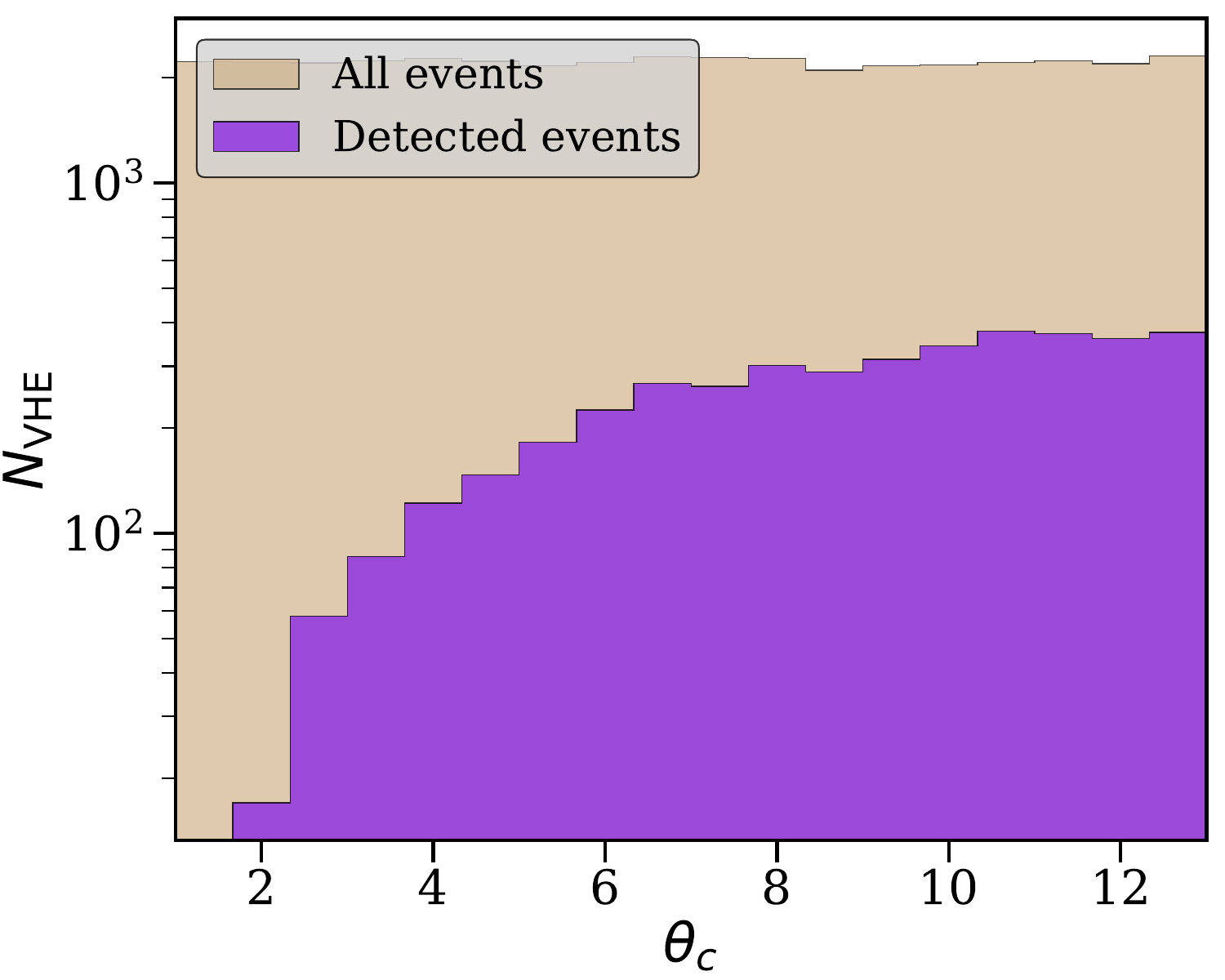}
\includegraphics[width=0.32\linewidth]{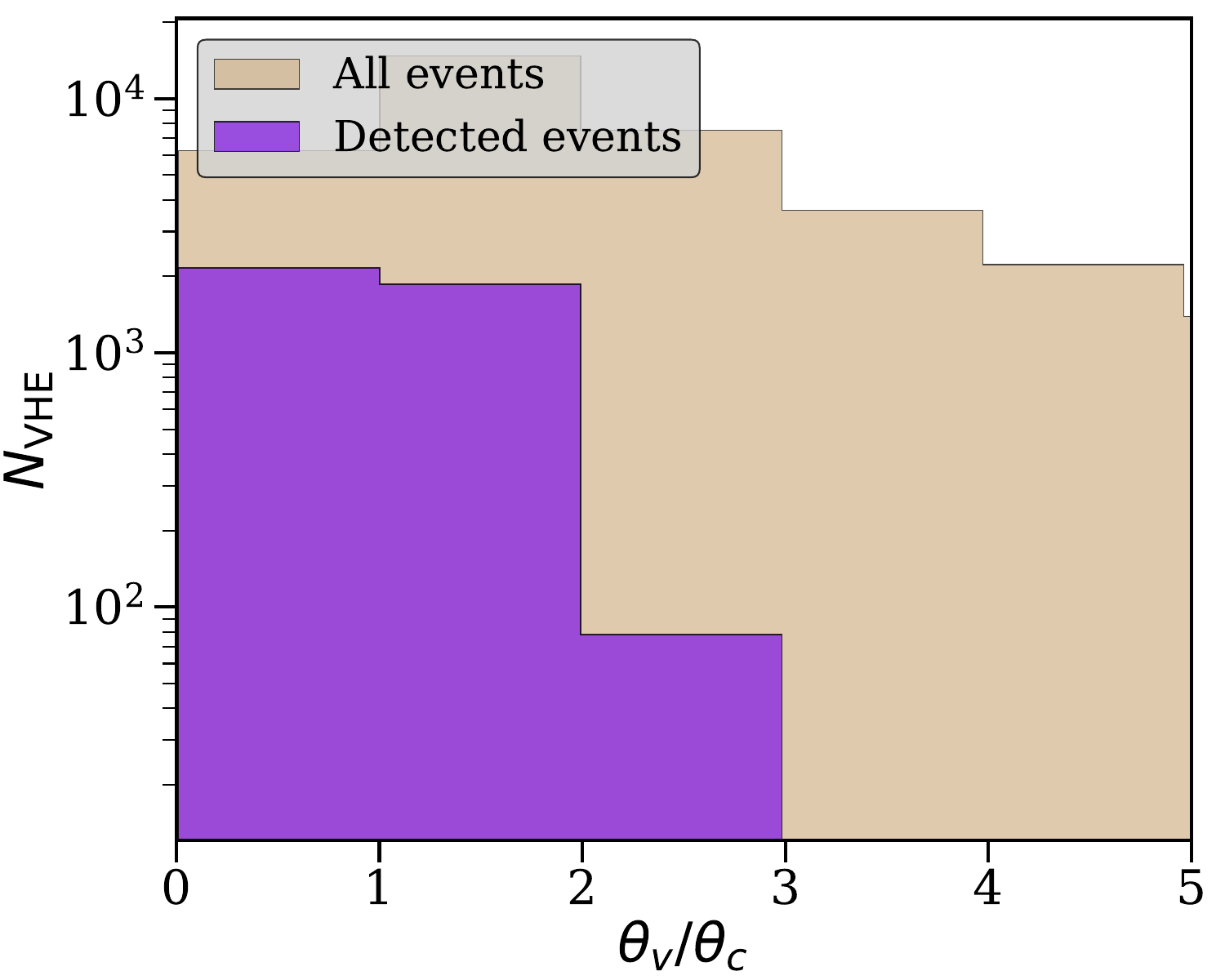}
\caption{Distribution of CTA-detected events compared to all simulated samples for parameter $\theta_v$, $\theta_c$ and $\theta_v/\theta_c$}
\label{fig:Hist_thv_thc}
\end{figure*}

Based on this prior distribution, out of the total sample, only $4101$ number of samples crossed the CTA sensitivity limit, underlining that TeV detections arise only from critical regions of parameter space, highlighting the strong parameter dependence. High kinetic energy $E_k$, wind number density $A_{\star}$ and moderate Lorentz factor $\eta_{c}$ increase the chance of detection (Figure~\ref{fig:Hist_Ek_A_etac_dL}). Since we have considered EBL attenuation, the detected events decline with increasing luminosity distance~\citep{mondal2025follow, mondal2023probing}. From Figure~\ref{fig:Hist_epse_epsB} we see, $\epsilon_{e}$ and $\epsilon_{B}$ play a decisive role in detection. Large $\epsilon_{e}$ enhances the synchrotron photon field, whereas very small $\epsilon_{B}$ reduces magnetic cooling, favouring efficient SSC upscattering into the TeV band. Similarly from Figure~\ref{fig:Hist_thv_thc}, the parameters deciding jet structure $\theta_{v}$ and $\theta_{c}$ strongly affect the detection. $\theta_{v}/\theta_{c}$ governs apparent brightness, thus detections are common for $\theta_{v}/\theta_{c}\lesssim 2$ (on-axis or near-core), decrease rapidly around $\theta_{v}/\theta_{c}\sim2$–$3$, and are almost absent beyond $\theta_{v}/\theta_{c}>3$. Together, these trends demonstrate that CTA detectability requires both favourable afterglow and microphysical parameters, along with favourable jet geometry.%, consistent with structured-jet interpretations of VHE-bright bursts such as GRB~221009A.

\section{Fitting VHE data of GRB 221009A in wind-driven medium}
\label{sec:fit_VHE_data}

In the previous section, we presented the theoretical framework for TeV emission from off–axis Gaussian structured jets in a wind medium. Here we apply that framework to \emph{light–curve} modeling of GRB~221009A across three bands—X–ray, GeV, and TeV regime. Our goal is to quantify the synchrotron and SSC–driven afterglow as it evolves in time, and to determine the model parameters that best reproduce the observed temporal evolution across these bands. To assess the detectability of the X–ray, GeV-TeV flux from GRB 221009A, we perform a comprehensive exploration of the afterglow parameter space by constraining the structured jet properties that favour VHE emission.

\subsection{Fitting Method}

\begin{figure}[ht]
    \centering
    \includegraphics[width=\linewidth]{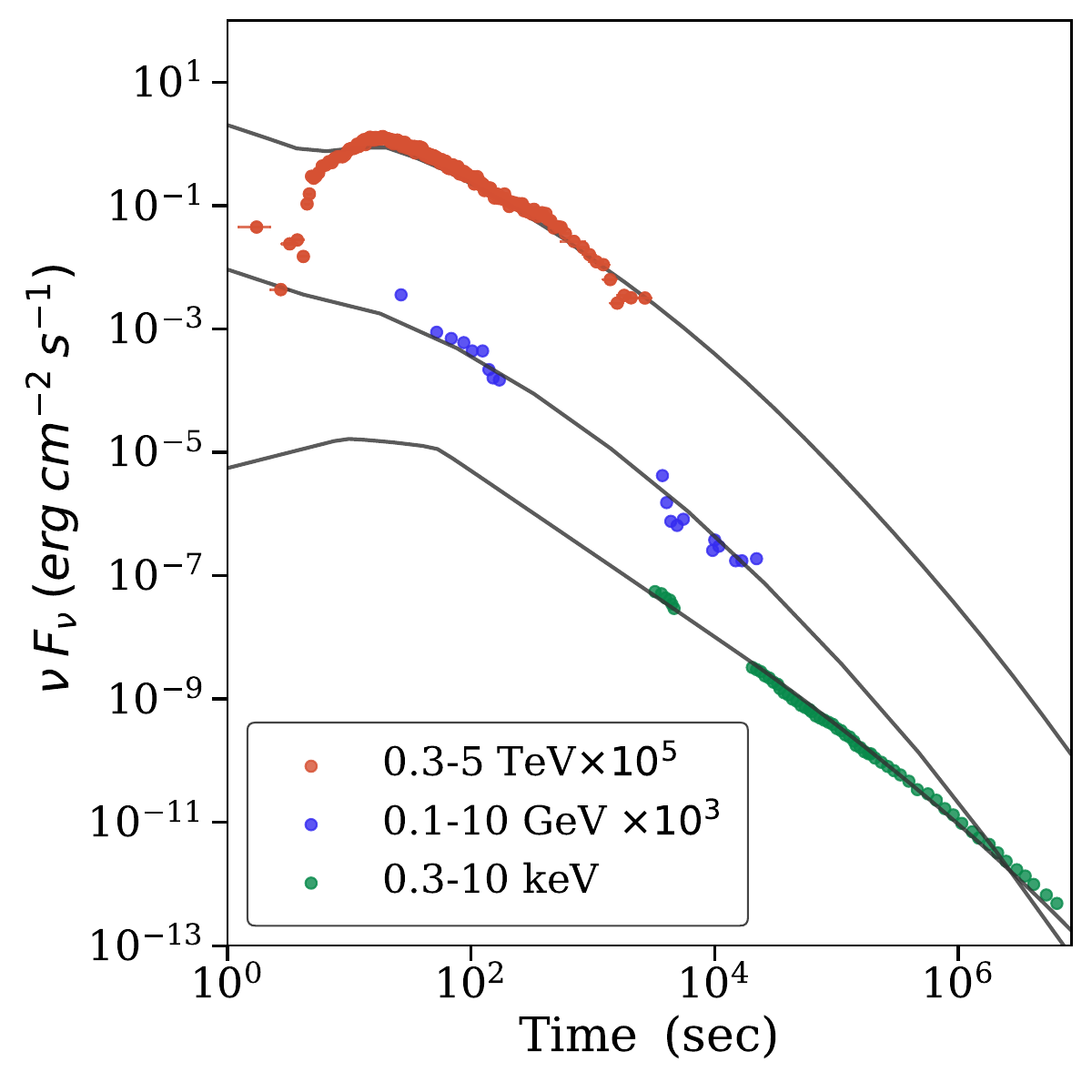}
    \caption{{Broadband afterglow light curves of GRB 221009A, with a Gaussian structured jet model in a wind-driven medium. Solid grey curves show Broadband afterglow light curves fitted with the wind–Gaussian structured-jet model. Data points includes: Red—0.3–5~TeV ($\times10^{5}$), violetblue—0.1–10~GeV ($\times10^{3}$), green—0.3–10~keV.}}
    \label{fig:wind_gauss_fit}
\end{figure}

\begin{figure}[ht]
    \centering
    \includegraphics[width=\linewidth]{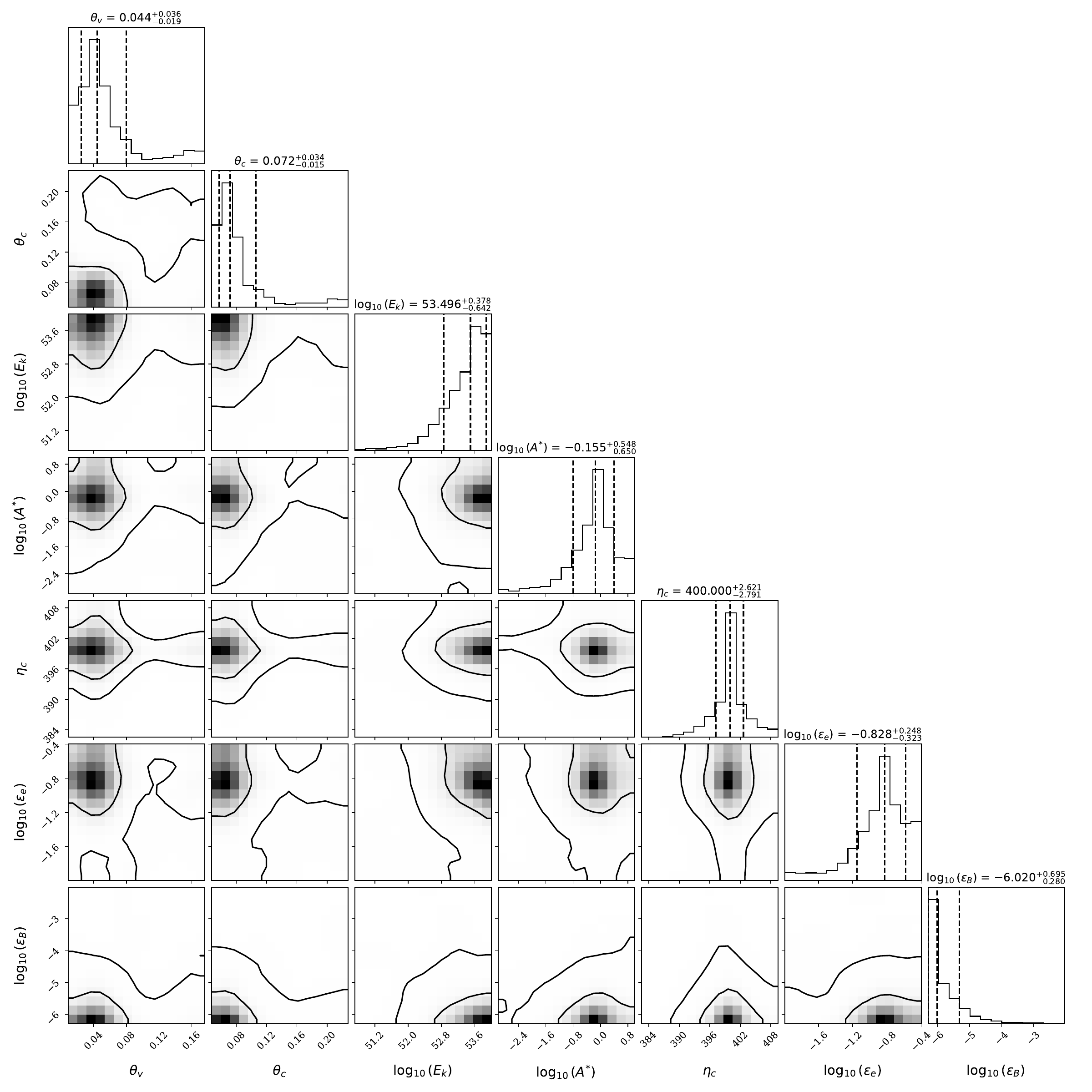}
    \caption{Corner plot showing posterior distributions of afterglow model parameters for GRB 221009A. All the contours enclosing median values and $\pm1 \sigma$ uncertainties. Thin reference lines denote the truth values.}
    \label{fig:corner}
\end{figure}

As outlined in Section~\ref{sec:model}, the peak flux and peak epoch of a GRB afterglow are jointly governed by the afterglow parameters and the jet geometry. We model GRB~221009A with a Gaussian structured jet expanding into a stellar wind, characterized by the \emph{nine} parameters, $\Theta_A=\{E_k,\, A_{\star},\, \eta_c, \, \theta_j,\, \theta_c,\, \epsilon_e,\, \epsilon_B,\, p,\, \theta_v\}$, and a luminosity distance fixed to $d_L=723.6~\mathrm{Mpc}$. For the light–curve fit (X–ray, GeV, and TeV), we vary seven parameters, while holding $\theta_j$ and $p$ fixed to $15^{\circ}$ and $2.5$ respectively, as $\theta_j$ presents the outer jet boundary, beyond which the energy drops rapidly. The Swift/XRT and LHAASO light–curve data used in this work are taken from \citet{ren2024jet, foffano2024theoretical} and \citet{lhaaso2023tera}, respectively.

We infer the posterior distributions of the model parameters with a Bayesian Markov Chain Monte Carlo (MCMC) analysis using the \emph{emcee}\footnote{\url{https://emcee.readthedocs.io/en/stable/}} \citep{foreman2013emcee} Python package. The fit is performed directly on the Swift/XRT (X–ray) and LHAASO (GeV–TeV) light curve data, adopting broad, physically motivated uniform priors for all seven parameters. Assuming independent Gaussian measurement errors, the likelihood is constructed from the reported flux uncertainties in each band, and instrumental systematic errors are not included. Thus, we efficiently explore the multidimensional parameter space and achieve robust posterior credible intervals for the wind-structured jet model of GRB~221009A.

\begin{table}[ht!]
\caption{Best-fit afterglow parameters from fits to the VHE light curves of GRB 221009A. Uncertainties are $1\sigma$ credible ranges from the MCMC posterior distributions.}
\label{tab:afterglow_params}
\vspace{2mm} 
\centering
\begin{tabular}{cc}
\hline
\hline
\textbf{Parameter}  & \textbf{Range}     \\ \hline
$\log_{10}E_k$    & $53.49^{+0.38}_{-0.64}$          \\[1.5mm]
$\log_{10}A_{\star}$ & $-0.15^{+0.55}_{-0.65}$            \\[1.5mm]
$\eta_c$                    & $399.99^{+2.62}_{-2.79}$            \\[1.5mm]
$\log_{10}\epsilon_{e}$     & $-0.83^{+0.25}_{-0.32}$          \\[1.5mm]
$\log_{10}\epsilon_{B}$     & $-6.02^{+0.069}_{-0.28}$          \\[1.5mm]
$\theta_{v}$ (deg)  & $2.52^{\circ}\,^{+2.06}_{-1.10}$ \\[1.5mm]
$\theta_{c}$ (deg) & $4.12^{\circ}\,^{+1.95}_{-0.84}$  \\
\hline
\end{tabular}
\end{table}

To achieve the temporal fit of GRB~221009A, we first sampled the parameter posteriors against the $0.1$–$10~\mathrm{GeV}$ light curve, using \textsc{emcee} with 32 walkers with 2000 iterations. The maximum likelihood (best–fit) parameters and their $1\sigma$ credible intervals are listed in Table~\ref{tab:afterglow_params}, and the joint posteriors are shown in the corner plot (Figure~\ref{fig:corner}). From the $1\sigma$ ranges of $E_k$, $\theta_c$, and $\theta_v$ we also infer the corresponding critical energy $\varepsilon_c$. The resulting model reproduces the observed X–ray, GeV, and TeV light curves of GRB~221009A with good agreement.

\subsection{Temporal modeling of the GRB~221009A afterglow}\label{sec:temporal}

In this work, we test the wind Gaussian structured-jet framework using only the multi-band light curves of GRB~221009A. Specifically, we perform a joint temporal fit to the Swift/XRT X-ray data ($0.3-10$~keV), LHAASO GeV data ($0.1-10$~keV), and TeV data ($0.3–5$~TeV). The model computes synchrotron and synchrotron self-Compton (SSC) emission from a relativistic outflow with a Gaussian angular structure that decelerates in a stellar-wind medium (see section~\ref{sec:model}). For the TeV band, we include Klein–Nishina (KN) corrections to the inverse-Compton cross section, and attenuation by the extragalactic background light (EBL). All light-curve fits are obtained with an MCMC sampler; quoted parameter uncertainties correspond to the $1\sigma$ credible intervals from the posterior distributions (see Table~\ref{tab:afterglow_params}).

The resulting very high energy (VHE) light curves display the characteristic two-phase evolution of a decelerating blast wave: a gradual rise to peak followed by a steeper decay. Within the structured-jet geometry, the temporal behaviour is regulated by both dynamics (kinetic energy and ambient density) and geometry (viewing angle $\theta_v$ relative to the core width $\theta_c$). A mildly off-axis configuration ($\theta_v/\theta_c \lesssim 2$) yields enhanced early-time VHE emission due to beaming from the energetic jet core, while larger $\theta_v/\theta_c$ values suppress and delay the peak. The microphysical parameters also plays a key role: fits consistently favour $\epsilon_e \gg \epsilon_B$, a regime that strengthens SSC relative to synchrotron at high energies and is required to reproduce the observed VHE SSC flux.

Figure~\ref{fig:wind_gauss_fit} summaries the broadband temporal fits across X-ray, GeV, and TeV bands. The model captures the rise–peak–decay characteristics and the relative timing offsets between bands that arise from the interplay of jet structure, blast-wave deceleration in a wind profile, and KN/EBL effects at the highest energies. The best-fit parameters inferred from the light curves are listed in Table~\ref{tab:afterglow_params}.

In some recent works, one by \citet{gill2023grb} has explained mostly X-ray to radio light curves using a structured jet power-law model. However, in comparison with the best fit model parameters, our model fit GeV-TeV data with relatively higher values of $E_{k}$, $A_{\star}$ and $\epsilon_{e}$ and lower values of $\epsilon_{B}$, whereas jet structure parameter in their case restrict to narrow jet core ($\sim1.2^{\circ}$) and small viewing angle($\sim1.14^{\circ}$), in our case observed data well fit with jet structure parameter $\theta_{c}\sim4.12^{\circ}$ and $\theta_{v}\sim2.52^{\circ}$. Further \citet{o2023structured} also explains X-ray and OIR late-time light curve features powered by a structured jet with a shallow angular profile. Their best-fit microphysical parameters well resemble to our case, except for jet total kinetic energy and the number density of the medium. Recently \citet{ren2024jet}, have used a structured jet model with a top-hat inner core and a power-law outer wing structure to explain multi-wavelength data of GRB 2210009A. 

\section{Summary and conclusions}
\label{sec:conclusions}
In this work, we present an afterglow framework in which a Gaussian structured jet propagates into a stellar–wind medium ($\rho\propto r^{-2}$), and study how jet afterglow parameters affect the CTA detectability. We then apply the model to the broadband (X–ray, GeV, TeV) afterglow of GRB~221009A. Through our Gaussian Structured jet model, we compute synchrotron and SSC radiation at VHE with Klein–Nishina (KN) suppression, and extragalactic background light (EBL) absorption. Within this wind-driven medium with Gaussian structured jet, we primarily explored the roles of geometry (the viewing ratio $\theta_v/\theta_c$), energetics (core energy and initial Lorentz factor), and microphysics $(\epsilon_e,\epsilon_B,p)$ in shaping the observed light curves.

In section~\ref{sec:model}, we explained the Gaussian structured jet model in a wind medium in detail to explain the synchrotron and SSC emission. Also, we discuss the effect of a wind-driven medium on magnetic field strength, peak flux, and peak time. Section~\ref{sec:param_charc} explores the effect of jet energetics, density of circumburst wind medium, microphysical parameters and jet geometry in shaping the VHE SSC light curves. We observe that the peak time and peak flux of the VHE light curve are governed by geometry through the ratio $\theta_v/\theta_c$, with on-axis views producing early, bright peaks and off-axis views producing delayed, suppressed emission. The Gaussian profile provides a smooth transition with better control over post-peak decay slopes across viewing angles. Our Gaussian structured jet model for the VHE regime favours $\epsilon_e\gg\epsilon_B$, consistent with SSC dominance at TeV energies, while the observed temporal evolution is strongly governed by the distinct values of microphysical parameters. These results have direct implications for VHE detectability. In the wind medium and Gaussian jet scenario, slight changes in $\theta_v/\theta_c$, $E_{k}$, and $A_\star$ can modify the TeV flux by orders of magnitude. 

We have further implemented the CTA detectability criterion based on the Gaussian structured jet wind model in section~\ref{sec:detect_TeV_events}. The result showcase that, higher value of $E_k$, $A_\star$, and $\epsilon_e$ favours the detection. Whereas the $\eta_c$ and $\epsilon_B$ have a moderate effect on SSC emission during the CTA detection. Further, the jet structure parameters depict that, higher value of $\theta_v$ decreases the number of detected events. Although with the increase of $\theta_c$, the overall detection increases. $\theta_{v}/\theta_{c}\lesssim 2$ provide the most favourable detection criteria for CTA. The afterglow model parameters for CTA detected cases in wind-driven medium follow a similar trend to those reported in~\citet{mondal2025follow}.

Finally, in section~\ref{sec:fit_VHE_data}, we have fitted the multi-band data (Xray-GeV-TeV) of GRB221009A implementing our model. The fit to GRB~221009A favours a mildly off-axis, near-core geometry in a wind medium, where we incorporate the effect of KN suppression and EBL attenuation. All light-curve fits are obtained with an MCMC sampler with the parameter uncertainties corresponding to the $1\sigma$ credible intervals from the posterior distributions. Within the structured-jet configuration, the temporal behaviour of GRB~221009A is jointly governed by jet total kinetic energy, wind density parameter, microphysical parameters and jet geometry ratio $\theta_v/\theta_c$ that controls relativistic beaming, and thus governs peak flux and peak time. 

In summary, a Gaussian structured jet in a wind medium provides a physically motivated, predictive description of GRB~221009A’s broadband afterglow and a general pathway to interpret VHE emission in long GRBs. By unifying jet geometry, jet kinematics, wind density, microphysical parameters and very high energy radiative processes within the structured jet SSC framework, we clarify why TeV detections with CTA are rare yet achievable, and we describe the conditions under which CTA is most likely to observe them.

\section{Acknowledgements}
T.M. and S.C. acknowledge the Prime Minister’s Research Fellowship (PMRF). R.L. acknowledges the support of the Anusandhan National Research Foundation (ANRF) grant CRG/2022/00803. T.M. expresses sincere gratitude to Prof. Sonjoy Majumder of the Department of Physics, IIT Kharagpur, for his invaluable guidance and encouragement. The authors also gratefully acknowledge the use of the Paramshakti Supercomputing Facility at IIT Kharagpur, established under the National Supercomputing Mission, Government of India, for providing the high-performance computational resources essential to this work.

\bibliographystyle{elsarticle-harv} 
\bibliography{GRB_wind}
\end{document}